\documentclass[useAMS,usenatbib]{mn2e}

\usepackage{natbib}
\usepackage{psfig}
\usepackage{graphicx}
\usepackage{amsmath}
\usepackage{amssymb}
\usepackage{color}

\newcommand{\LCDM}{$\Lambda$CDM}
\newcommand{\msun}{\mbox{${\rm M}_{\odot}$}}

\newcommand{\fred}{\mbox{$f_{\rm red}$}}
\newcommand{\fpass}{\mbox{$f_{\rm passive}$}}
\newcommand{\fearly}{\mbox{$f_{\rm early}$}}
\newcommand{\mgal}{\mbox{$M_{\rm gal}$}}
\newcommand{\mhalo}{\mbox{$M_{\rm halo}$}}

\def\lesssim{\lower.5ex\hbox{$\; \buildrel < \over \sim \;$}}
\def\gtrsim{\lower.5ex\hbox{$\; \buildrel > \over \sim \;$}}

\title[Quenching: internal properties and environment]
{The Correlation of Star Formation Quenching with Internal Galaxy Properties and Environment}

\author[Taysun Kimm et al.]{
\parbox[t]{\textwidth}{
Taysun Kimm$^{1}$,
Rachel S. Somerville$^{2,3}$, 
Sukyoung K. Yi$^{1}$\thanks{e-mail: yi@yonsei.ac.kr},
Frank C. van den Bosch$^{3}$, 
Samir Salim$^{4}$, 
Fabio Fontanot$^{3}$, 
Pierluigi Monaco$^{5,6}$, 
Houjun Mo$^{7}$,
Anna Pasquali$^{3}$,
R. M. Rich$^{8}$,
Xiaohu Yang$^{9}$
}
\vspace*{6pt} \\
$^1$ Department of Astronomy, Yonsei University, Seoul 120-749, Korea\\
$^2$ Space Telescope Science Institute, 3700 San Martin Dr., Baltimore, MD 21218, USA\\
$^3$ Max-Planck-Institut f\"{u}r Astronomie, K\"{o}nigstuhl 17, Heidelberg 
D-69117, Germany\\
$^4$ National Optical Astronomy Observatory, 950 North Cherry Ave., Tucson AZ 85719, USA\\
$^5$ Dipartimento di Astronomia, Universit\`a di Trieste, via Tiepolo 11, I-34143 Trieste, Italy \\
$^6$ INAF-Osservatorio Astronomico, Via Tiepolo 11, I-34143 Trieste, Italy \\
$^7$ Department of Astronomy, University of Massachusetts, Amherst MA01003-9305, USA \\
$^8$ Department of Physics and Astronomy, University of California, Los Angeles, CA 90095, USA\\
$^9$ Shanghai Astronomical Observatory, the Partner Group of MPA, Nandan Road 80, Shanghai 200030, China\\
\\
}

\begin{document}
\date{\today}
\maketitle
\label{firstpage}

\begin{abstract}
We investigate the correlation of star formation quenching with
internal galaxy properties and large scale environment (halo mass) in
empirical data and theoretical models.  We make use of the halo-based
Group Catalog of Yang and collaborators, which is based on the Sloan
Digital Sky Survey. Data from the Galaxy Evolution Explorer (GALEX)
are also used to extract the recent star formation rate.  In order to
investigate the environmental effects, we examine the properties of
``central'' and ``satellite'' galaxies separately.  For central
galaxies, we are unable to conclude whether star formation quenching is
primarily connected with halo mass or stellar mass, because these two
quantities are themselves strongly correlated. For satellite
galaxies, a nearly equally strong dependence on halo mass and stellar
mass is seen.  We make the same comparison for five different
semi-analytic models based on three independently developed codes. We
find that the models with AGN feedback reproduce reasonably well the
dependence of the fraction of central red and passive galaxies on halo
mass and stellar mass.  However, for satellite galaxies, the same
models badly overproduce the fraction of red/passive galaxies and do
not reproduce the empirical trends with stellar mass or halo mass.
This {\em satellite overquenching problem} is caused by
the too-rapid stripping of the satellites' hot gas halos, which leads
to rapid strangulation of star formation.
\end{abstract}

\section{Introduction}
\label{sec:intro}

Galaxies may be broadly divided into two categories: those that are
forming stars fairly rapidly relative to their past averaged star
formation rate (SFR), and those that show little recent star formation
relative to their past average. It is well known that, at least at low
redshift ($z\lesssim 1$), the former type tend to have blue colours
and to be morphologically disk dominated, while the latter tend to
have red colours and to be morphologically early type or spheroid
dominated. Galaxies exhibit colour bimodalities throughout a wide
range of cosmic history
\citep{strateva01,bell04,baldry04,balogh04,blanton05b}.  One of the
fundamental questions in galaxy formation is: what are the main
physical forces that regulate and in some cases quench star formation?
Are these processes more closely correlated with {\em internal galaxy
properties} such as mass or luminosity, or {\em large scale
environment} (sometimes referred to as ``Nature or Nurture'')?

There have been many studies on the impact of environment on galaxy
properties. For example, \citet{davis76} and many
others found that early-type galaxies are more strongly clustered
than the late-types, and \citet{dressler80} systematically
demonstrated that the fraction of elliptical and S0 galaxies is
higher in denser environments. Similarly, it is well known that
galaxies in dense environments tend to be red and to have depressed
star formation rates \citep{hashimoto98,lewis02,gomez03,tanaka04,balogh04,
kauffmann04,christ05,poggianti06}.  There are several physical
processes associated with environment that may play a role in galaxy
transformation. Major (near equal-mass) mergers can transform spiral
galaxies into ellipticals (\citealt{toomre72}; cf.
\citealt{barnes02}), and may also quench future star formation by
ejecting the ISM from the galaxy via starburst, AGN, or shock-driven
winds \citep{cox04,springel05,murray05}. In rich clusters, where the
probability of merging is suppressed because of the large relative
velocities of galaxies, galaxy ``harrassment'' (rapid encounters or
fly-bys) may cause a less dramatic form of transformation by heating
disks, perhaps causing the formation of a bar and growth of a
spheroidal component \citep{moore98}.  Also, cold gas can be stripped
out of the galaxy both by tidal forces due to the background dark
matter dominated potential of the cluster, and due to ram pressure
stripping by the intracluster medium \citep{gunn72,abadi99,
  quilis00,chung07}. Similarly, the hot halo that provides future fuel
for cooling and star formation may be efficiently stripped in dense
environments, thus quenching further star formation by ``starvation''
or ``strangulation'' \citep{larson80,balogh00,bekki02}. 

However, studies by \citet{vdb08a} and \citet{tanaka04} 
suggest that processes specific to clusters
(e.g. ram-pressure stripping) are not the main mechanisms for
quenching star-formation activity. Similar results were also found
at higher redshift \citep[e.g.,][]{cooper06}.  Moreover, both the
morphological and spectrophotometric characteristics of galaxies are
also known to be strongly correlated with their internal properties,
such as luminosity, mass, and internal velocity
\citep{roberts94,kauffmann03}.  More massive or luminous galaxies are
more likely to be spheroid dominated, to be red, and to have old
stellar populations and little recent star formation. Indeed,
\citet{kauffmann03} showed that galaxies appear to make a transition
in all of these properties above a critical stellar mass of $\sim 3
\times 10^{10} M_{\odot}$.

Again, there are various physical processes that one might expect to
imprint this kind of dependence on internal properties. Supernova
feedback has long been invoked as a mechanism that could heat and
drive gas out of galaxies \citep{larson74,dekel86}, and is
expected to be more effective in low-mass galaxies. There is also
mounting observational evidence that AGN are associated with the
quenching of star formation \citep{schawinski06,schawinski07,salim07}. 
AGN feedback is expected to have more impact on massive galaxies,
which host larger mass black holes \citep[e.g.][]{silk98}.

The emerging picture is that AGN seem to have two modes of fueling and
also to couple to their surroundings in different ways. ``Bright
mode'' AGN are associated with high (near Eddington) fueling rates,
and observationally with classical X-ray or optically bright
quasars. The radiation emitted by these objects can couple with the
cold gas in the galaxy, perhaps driving powerful winds that can drive
the gas out of the galaxy and quench star formation
\citep{murray05,dimatteo05,monaco05}.

In contrast, many massive galaxies seem to contain AGN which are
accreting at a small fraction of their Eddington rate, and which
typically do not show classical quasar-like signatures such as bright
X-ray radiation or broad emission lines. However, these objects are
associated with the efficient production of radio jets, which may be
able to couple to and heat the hot gas in the galaxy's halo
\citep[e.g.][]{mcnamara07}. These objects are often referred to as
``radio mode'' AGN \citep{croton06}. 
This process is expected to be a function both of
internal properties and environment: the bigger the black hole, the
more energy can be tapped (and black hole mass is of course correlated with
galaxy mass), and empirically it is known that the fraction of
radio-detected galaxies increases strongly with stellar mass \citep{best05} 
and halo mass \citep{pasquali08}. 
But as well, the radio jets must have a ``working surface''
and therefore can only be effective in environments that can support a
quasi-hydrostatic hot gas halo, such as groups and clusters. Galaxies
in smaller mass halos ($\mhalo \lesssim 10^{12} \msun$) likely accrete
most of their gas in a ``cold flow'', and never form a hot halo
\citep{birnboim03,dekel06,keres05}.

A great deal of recent progress has been made towards developing a
comprehensive theory of galaxy formation.
The semi-analytic approach, although it has its limitations, is a
powerful and flexible tool for exploring detailed predictions based on
this theor
\citep[e.g.,][]{wf91,kauffmann93,kauffmann99,sp99,cole00,spf01,
  springel01,benson03,hatton03,khochfar03,khochfar05,kang05}.
Several groups have now implemented one or both modes of black hole
growth and AGN feedback into their semi-analytic models 
\citep[e.g.,][]{croton06,bower06,monaco07,s08}.
There seems to be consensus that including
these new processes leads to greatly improved agreement with key
observations such as galaxy luminosity or mass functions, and the
galaxy colour-magnitude distribution or stellar mass vs. specific star
formation distribution. However, this necessitates including several
new recipes and parameters associated with the poorly understood
physics of black hole growth and AGN feedback. Each model contains somewhat
different parameterizations and treatments of these processes, yet
they all produce similar results for global quantities such as the
galaxy luminosity function and colour-magnitude distribution, no doubt
in part because these observations were ``targets'' that the modellers
were trying to reproduce. One goal of our work here is to determine
whether breaking down the fraction of quenched galaxies in the dual
space of internal galaxy properties and environment can discriminate
between these different treatments of AGN feedback.

The large and homegeneous databases provided by modern surveys, such
as SDSS and GALEX, finally allow a statistically significant
investigation of different sub-populations of galaxies. However, an
obvious question that arises in any study of galaxy environment is
exactly how to measure and characterize environment in observational
samples. Clearly, it is desirable to span as broad a range of
environments as possible, from isolated field galaxies to groups to
rich clusters. The majority of the studies in the literature have
parameterised environment in terms of the number of
galaxies, either within a fixed metric aperture, or by the distance of
the $n$th-nearest galaxy, where $n$ is typically in the range 3--10. 
Although these indicators are straightforward to measure, they
are not straightforward to interpret in physical terms or to compare
with theoretical models (see discussions in \citealt{kauffmann04} and
\citealt{weinmann06a}). An alternate approach is to use a galaxy group
catalogue, in which galaxies in an observational catalog are not only
grouped together into putatively gravitationally bound structures, but
also the total mass of their associated dark matter halo is estimated
(Yang et al. 2007). Thus, the relationship between galaxy properties and dark
matter halo properties can be studied directly. Another advantage of
this method is that galaxies can be separated into ``central'' and
``satellite'' populations. Most of the environment-related
tranformation mechanisms described above (such as stripping) are
expected to work only on satellite galaxies, so this offers a way to
separate out the effects of different physical processes.

Here we make use of a large galaxy group catalogue constructed from 
the SDSS using the halo-based galaxy group finder developed by 
\citet{yang05}. These catalogues have already been used for 
several studies regarding the environment dependence of galaxy properties. 
\citet{weinmann06a}, using the version based on the SDSS DR2, studied 
the correlations between colours, specific star formation rate and halo mass. 
Splitting the galaxy population into early and late types, based on their 
colours and specific star formation rates, they found that, at a fixed 
luminosity, the late (early) type fraction of  galaxies increases 
(decreases) with decreasing halo mass. Using the much larger galaxy 
group  catalogue of Yang et al. (2007; hereafter Y07), based on the 
SDSS DR4, \citet{vdb08a} showed that, on average, satellite 
galaxies are redder and more concentrated than central galaxies of 
the same stellar mass. They also found that the colour and concentration 
differences of central-satellite pairs matched in stellar mass are 
completely independent of the mass of the host halo of the satellite 
galaxy. This indicates that satellite-specific transformation mechanisms 
are equally efficient in host haloes of all masses and rules against
satellite transformation mechanisms that are thought to operate only in 
very massive haloes. Further support for this was provided by \citet{vdb08b} 
and \citet{pasquali08}, who showed that, at fixed 
stellar mass, the average colours and concentrations, as well as the 
occurrence of star formation and AGN activity, reveal only a very weak 
dependence on halo mass (but see \citealt{weinmann08}).

\citet{weinmann06b} compared the fractions of red and blue
galaxies in the SDSS group catalogue of \citet{weinmann06a} with
the semi-analytic model of galaxy formation of \citet{croton06}.
Although this model accurately fits the global statistics of the
galaxy population, the model predicts a red fraction of satellites
that is much higher than observed 
\citep[see also][]{baldry06,coil08}.

In this paper we extend the study of \citet{weinmann06b} using the
much larger galaxy group catalogue of Y07 and a larger suite of
semi-analytic models.  Another new aspect of our work here is that we
augment the SDSS-based galaxy properties in the group catalogue with
information derived from the GALEX-SDSS matched sample of Salim et
al. (2007; hereafter S07). The S07 analysis provides complementary
quantities such as stellar masses and star formation rates based on
the two GALEX UV bands plus five-band SDSS photometry. The UV provides
a much more sensitive probe of the recent star formation history of a
galaxy than optical colours alone, which reflect the highly degenerate
effects of stellar populations, metallicity, and dust extinction
\citep{yi05,kaviraj07}.  In addition, the UV can provide reliable
measures of SFR for galaxies with weak or undetected emission lines.

The goal of this paper is to investigate the dependence of the
fraction of quenched galaxies on galaxy properties and DM halo mass in
both the observational group catalogue and in several different
semi-analytic galaxy formation models. Another new aspect of our study
with respect to previous comparisons \citep[e.g.][]{weinmann06b} is
that we compare with the results from three independently developed
semi-analytic codes, and for one of the codes, we examine three
different variants with different physical ingredients.  In this way,
we investigate how these empirical results can constrain the input
physics in these kinds of models. An outline of the rest of our paper
is as follows: in \S 2, we describe the empirical data sets and
the group catalogues used in our study; in \S 3 we describe the
theoretical models; in \S 4 we present the results of our comparisons
between the empirical data and the models; we discuss our results
and conclude in \S 5.

\section{The Data Sample}
\label{sec:data}

\subsection{The SDSS Group Catalog}
\label{sec:data:group}

We make use of the Galaxy Group Catalog of Yang et al. (2007; Y07).
The catalogue was constructed by applying the halo-based group finder
of \citet{yang05} to the New York University Value-Added Galaxy
Catalog (NYU-VAGC, \citealt{blanton05a}), which is based on the Sloan
Digital Sky Survey (SDSS) Data Release 4 (DR4; \citealt{adelman06}). 
From the Main Galaxy Sample, Y07 selected galaxies with
extinction-corrected $r$-band apparent magnitude brighter than $r=17.77$,
within a redshift range $0.01 < z < 0.2$, and with a redshift
completeness $\mathcal{C}>0.7$. They augmented this sample with 7091
galaxies with $0.01 <z< 0.2$ with redshifts from alternate
sources. The resulting sample (Sample II of Y07) has a mean redshift
of $z\sim0.1$ and a total sky coverage of 4,514 deg$^2$, and contains
369,447 galaxies. In this paper, we refer to this as ``the SDSS
sample''.

All absolute magnitudes are k and evolution (e) corrected to the
$z=0.1$ rest-frame, as described by \citet{blanton03}. Stellar masses
are computed using the relation between rest-frame optical colour and
mass-to-light ratio of \citet{bell03}, as specified in Y07, assuming a
Kroupa IMF.  Our sample is not volume limited, and we therefore
attempt to correct for the resulting Malmquist bias by weighting each
galaxy by a standard $V_{\rm max}$ correction \citep{schmidt68}.

The group finder first identifies a potential group centre by the
friends-of-friends algorithm \citep{davis85}, and then computes the
characteristic luminosity of each tentative group. The characteristic
group luminosity is defined as the (incompleteness corrected; see Y07)
combined luminosity of all group members with $^{0.1}M_r \leq -19.5 +
5\log h$. The characteristic group stellar mass is similarly the
incompleteness-corrected total stellar mass contributed by galaxies
with $^{0.1}M_r \leq -19.5 + 5\log h$. Then, the velocity dispersion
and the virial radius of the dark matter halo (DMH) associated with
each tentative group are estimated iteratively, assuming a constant
mass-to-light ratio as an initial guess. These two estimates can be
used for determining the spherical \citet{navarro97} (NFW)-type dark
matter profile.  Assuming a Gaussian probability distribution along
the redshift direction and a projected spherical NFW dark matter
profile for the perpendicular plane, group members are updated until
there is no change in their memberships.

Lastly, a DMH mass is assigned to each group, assuming a one-to-one
correlation between the characteristic stellar mass for each group and
the halo mass derived from a theoretical mass function 
\citep[e.g.][]{warren05}. We use the halo mass estimated from stellar mass,
as luminosity-based halo mass estimates may be biased differently for
groups with blue vs. red galaxies (Y07).  Y07 find that they can
assign group masses down to a lower limit of $M_{\rm halo} < 10^{11.6}
M_{\odot}/h$, although the completeness of the group catalogue begins
to drop below unity for halo masses less than $M_{\rm halo} \lesssim
10^{13} h^{-1} \msun$. The central galaxy in each group is identified
as the galaxy with the largest stellar mass.

There are 204,813 groups in the resulting group catalogue.
Note that we call all clusters ``groups'' regardless of their richness
or density, including groups that contain only a single member.  
For a more detailed description of the group finder, and the results
of extensive tests of its completeness, contamination, and purity, we
refer interested readers to \citet{yang07}.

\begin{table*}
\begin{center}
 \caption{Model descriptions}
\begin{tabular}{ c c c c c}\hline \hline
  Model & box size (Mpc/h) & Characteristics\\
  \hline
  S08 AGN-FB (fiducial) & 120 & Bright mode AGN-driven winds + Radio mode AGN heating\\
  S08 no AGN-FB & 120 & Control model without AGN feedback\\
  S08 HQ & 120 & cooling quenched according to the halo mass\\
  dL06 & 120 & Radio mode AGN heating \\
  {\sc morgana} & 144 & Bright mode AGN-driven winds + Radio mode AGN heating\\
\hline
 \end{tabular}
 \end{center}
\end{table*}

\subsection{GALEX}
\label{sec:data:galex}

The GALEX data provide near-UV (NUV, effective wavelength $\sim$
2271\AA) and far-UV (FUV, $\sim$ 1528\AA) band information \citep{martin05}.
Of 741 sq. deg. of GALEX unique imaging, 645
sq. deg. overlaps with the SDSS DR4 spectroscopic area. We make use of
the GALEX-SDSS matched sample constructed by Salim et al. (2007;
S07). S07 first define a SDSS parent sample in the GALEX overlap
region of objects spectroscopically classified as galaxies, and having
optical magnitude $14.5<r_{petro}<17.77$ and redshift
$0.005<z<0.22$. This sample contains 49,346 galaxies. For each of the
galaxies in the SDSS parent sample, S07 searched within a 4 arcsec
radius (corresponding to $\sim$ 7 kpc at $z\sim$0.1) for a match in
the GALEX source catalog.

Using a large library of model SEDs based on the \citet{bc03} 
population synthesis code, S07 then performed Bayesian SED fitting to
the 7-band photometry ($FUV, NUV, u, g, r, i, z$) to obtain estimates
of dust extinction, stellar mass, and star formation rate (SFR).  
These parameters were built from probability distribution functions, thus 
taking into account parameter degeneracies. 
The typical error in the specific SFR (obtained from the width of the SED
fitting probability distribution) is 0.2 dex (star-forming galaxies)
to 0.7 dex (passive galaxies).

Of the full S07 sample, we use 32,787 galaxies that are associated 
with SDSS galaxies in our group catalog, and exclude a small number of 
objects with very poor fits (see discussion in \S 4.3 in S07).
The redshift range for this final sample is $0.01\leq z \leq 0.2$ with
mean redshift $\sim$ 0.104, and the redshift distribution is very
similar to that of the parent SDSS sample.

\section{Theoretical Models}
\label{sec:models}

We adopt a semi-analytic approach to model galaxy formation within the
\LCDM\ picture. We make use of a total of 5 sets of models based on
different prescriptions for our comparison with the empirical
data.  Three of the models are constructed using the latest version of
the Somerville code (\citealt{sp99,spf01,s08}, hereafter S08).
The others are the Millennium models \citep{croton06,delucia06} and 
the MORGANA models \citep{monaco07,fontanot06,fontanot07}.

We describe the basic scheme in the Somerville code.  We use an N-body
simulation box to obtain the masses and positions of the ``root'' dark
matter halos, and compute the merger history for each halo using the
method of \citet{sk99}. Within each dark matter halo, gas cools via
atomic cooling \citep{wf91,sp99} and forms a rotationally supported
disk. The radial sizes of disks are computed using the model described
in \citet{somerville08a}, which accounts for the initial
Navarro-Frenk-White profiles of the DM halos and the ``adiabatic
contraction'' due to the self-gravity of the infalling baryons.  Cold
gas is turned into stars in the galactic disk following the
Schmidt-Kennicutt law \citep{kennicutt89,kennicutt98}, and gas with
surface density lower than a critical threshold density 
\citep[e.g.][]{martin01} does not form stars. Massive stars explode as
supernovae and reheat the cold gas. We trace chemical evolution using
a simple ``effective yield'' parameter. Each generation of stars
produces a fixed ``yield'' of metals, which are deposited in the cold
gas. This gas may then be ejected and mixed with the hot component by
supernova or AGN driven winds.

When a satellite is subsumed into a larger dark matter halo, it is
assumed that it immediately loses its hot gas halo, and thus does not
receive any new supply of cold gas.   The orbital decay and eventual
merging, due to dynamical friction, of satellite galaxies within dark
matter halos is tracked using a modified version of the
\citet{chandra43} formula \citep{boylan08}.  Mass loss and tidal
destruction are also accounted for, using a simplified version of the
approach presented in \citet{taylor04} and \citet{zentner03}. Galaxy
mergers trigger bursts of star formation, the efficiency and
timescales of which are modelled using results from hydrodynamic
simulations of galaxy-galaxy mergers.

The code also tracks the growth of black holes and the energy they
produce. Every top level DM halo is seeded with a black hole of $\sim
100 \msun$. Mergers trigger the ``bright mode'' black hole growth that
is associated with luminous quasars. Following every merger with mass
ratio greater than 1:10, the black hole grows at its Eddington rate
until it reaches a critical mass. The critical mass is that at which
the energy radiated by the black hole is sufficient to halt further
accretion, i.e., the black hole growth is self-regulated
\citep{hopkins07}.  Soon after reaching the critical mass, the black
hole enters a ``blowout'' phase, resulting in a decline in the
accretion rate. The associated radiation can also drive winds that
remove cold gas from the galaxy (see S08 for details).

The S08 models also incorporate ``radio mode'' feedback associated
with low-efficiency accretion. The accretion rate is computed using
the isothermal Bondi flow model of \citet{nulsen00}. In the presence
of a quasi-hydrostatic shock-heated gas halo, the energy from this
accretion is assumed to drive radio jets that can heat the gas and
partially or completely offset the cooling flow.

The resulting star formation and enrichment histories are convolved
with the stellar population models of \citet{bc03} to compute
magnitudes and colours. We have adopted a Chabrier IMF. The impact of
dust extinction is modelled using an analytic model, 
as in \citet{delucia07}.

The S08 {\em fiducial model} includes all of these mechanisms. We also
consider a ``no AGN feedback model'', which does not include either the
bright mode or radio mode AGN feedback mechanisms, and the ``Halo
Quenching'' model, in which cooling is shut off when a halo grows more
massive than $\sim 10^{12} M{\odot}/h^{-1}$ (see Table 1 for a summary
of all models). The halo quenching model is included as an
illustration of a quenching mechanism that has a simple dependence on
halo mass only. A similar model has been considered by
\citet{cattaneo06}. It is based on the ideas proposed by
\citet{birnboim03} and \citet{dekel06}, who suggest that whenever a
halo grows above this critical quenching mass, the gas is shock-heated
to near the virial temperature, and can be easily kept hot either by
an AGN or by other processes such as heating by gas clumps or orbiting
satellites \citep{dekel08,khochfar08}.

\begin{figure*}
\begin{center}
\includegraphics[width=16cm]{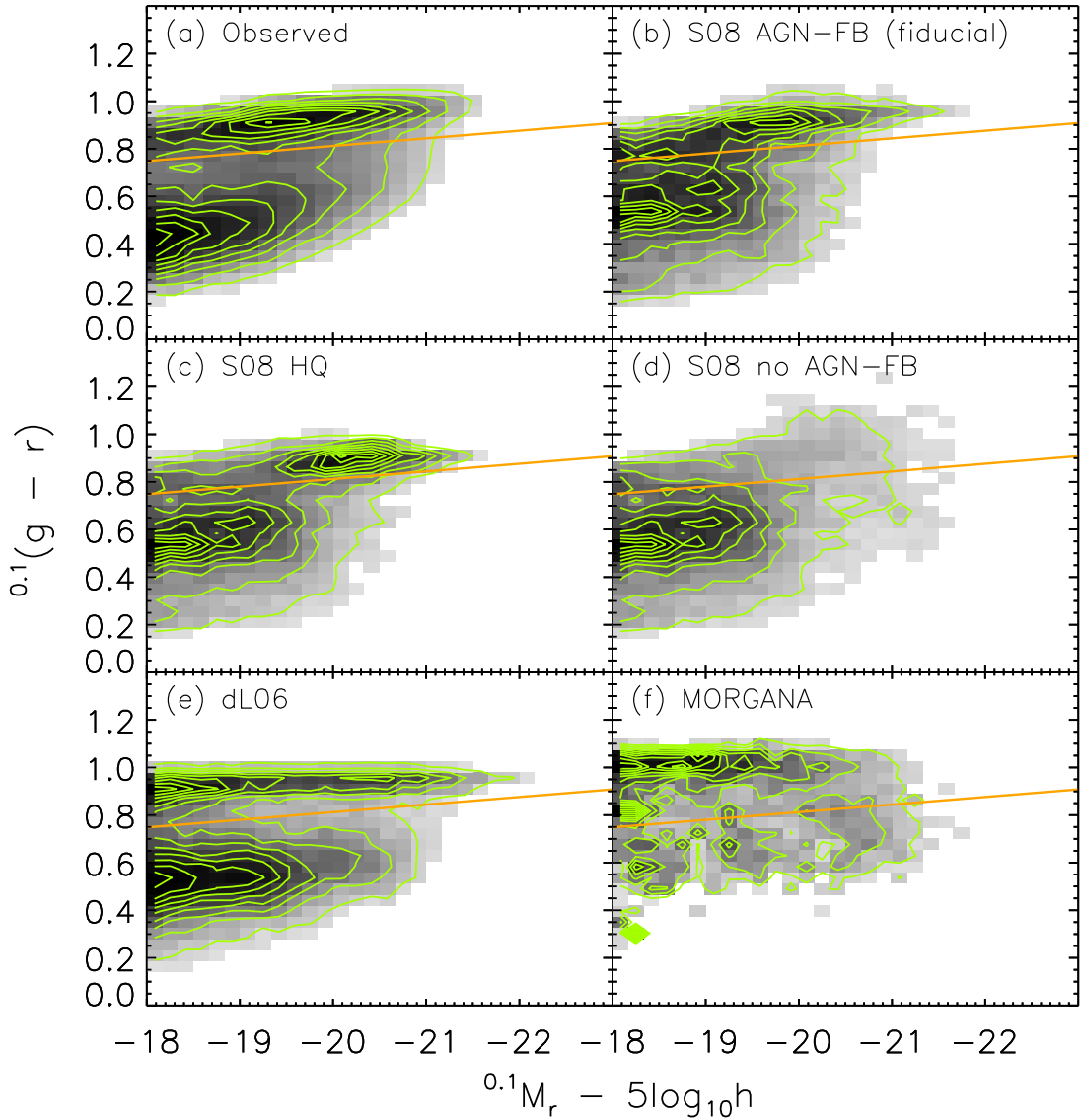}
\caption{The distribution of $^{0.1}r$-band absolute magnitude
  vs. $^{0.1}(g-r)$ colour.  Gray shading and green contours show the
  conditional probability P($^{0.1}M_r|^{0.1}(g-r)$). The orange line
  shows the demarcation line for the ``red'' and ``blue'' galaxy
  populations used in this study.  }
\label{fig:colourmag}
\end{center}
\end{figure*}

\begin{figure*}
\begin{center}
\includegraphics[width=16cm]{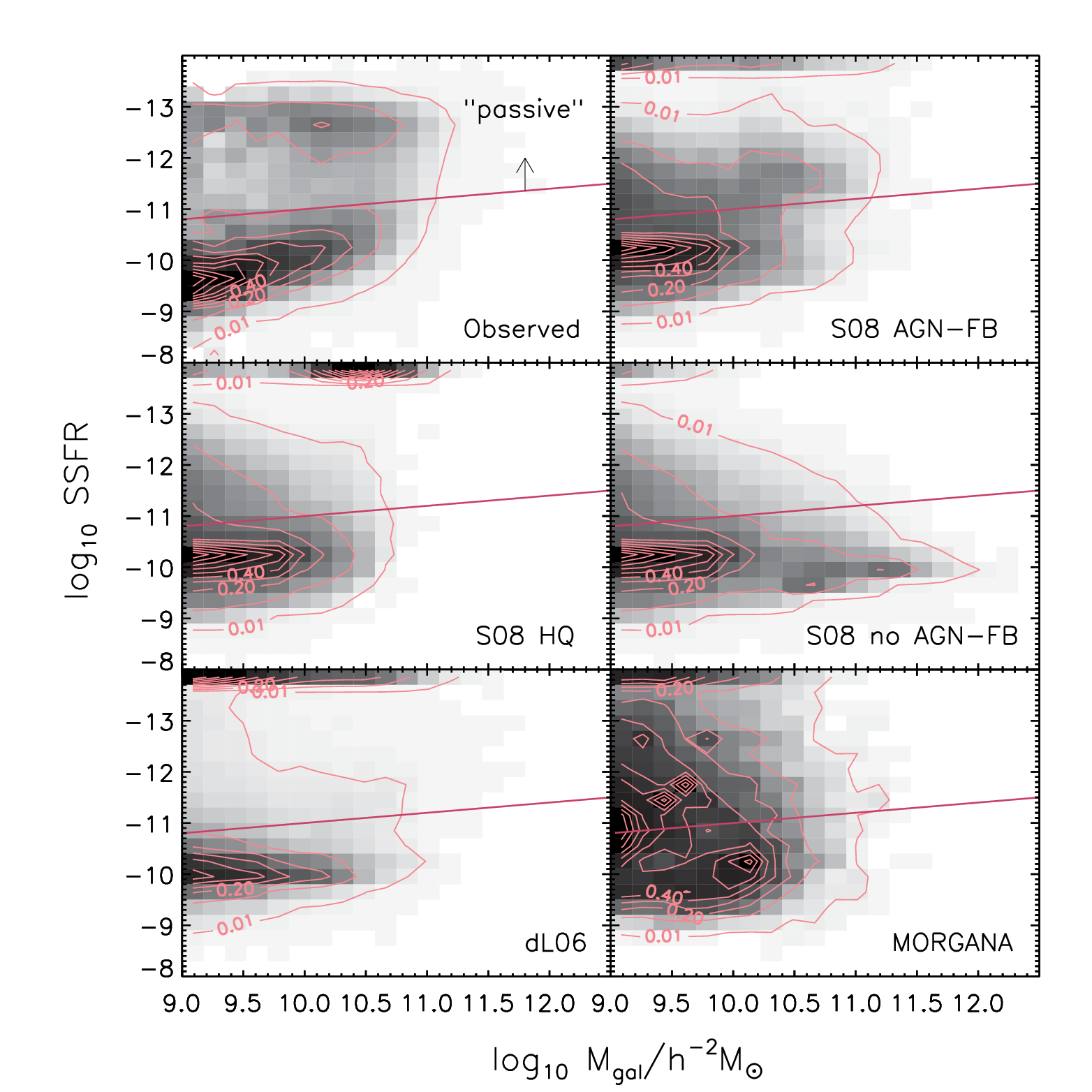}
\caption{Galaxy stellar mass vs. specific star formation rate.  Gray
  shading and pink contours show the conditional probability
  P($\mgal|SSFR$). The purple line shows the demarcation line between
  ``passive'' and ``active'' galaxies used in this study.
} \label{fig:ssfr}
\end{center}
\end{figure*}

We also consider two additional models from other groups in this
study.  The de Lucia et al. (2006, hereafter dL06) models contain
similar ingredients to the S08 models, with the following differences.
They are based on merger trees extracted from the Millennium N-body
simulations \citep{springel05nature}. 
Unlike in the models of S08, they do not include
the effects of adiabatic contraction and an NFW halo profile in their
estimates of galaxy sizes, which leads to a different evolution in the
star formation rates and gas fractions in their models. They use
somewhat different (though similar in spirit) recipes for star
formation and supernova feedback, and they use a different approach
for modelling black hole growth (though, like S08, they assume that
``bright mode'' black hole growth is triggered by mergers). They do not
include ``bright mode'' AGN feedback (AGN-driven winds), and their
treatment of the radio mode feedback is again similar in spirit but
different in detail from S08. Magnitudes and colours (including dust
extinction) are computed in a similar manner to S08, and use a
Chabrier IMF. We obtained the dL06 catalogs from the public Millennium
database (http://www.g-vo.org/Millennium).

We also consider the predictions of the semi-analytic model {\sc
  morgana} \citep{monaco07,fontanot06,fontanot07}.  {\sc morgana}
follows a scheme similar to S08 and dL06, but it includes a different
treatment for the thermal processes acting on baryonic gas. More
details on the updated version we use in this paper are presented in
Lo Faro et al. (2008, in prep).  The model is based on merger trees
obtained using the {\sc pinocchio} method \citep{monaco02}, similar to
those predicted by N-body simulations.  Gas cooling and infall follow
the prescription described and tested in \citet{viola08}, while star
formation and stellar feedback are then modeled as in
\citet{monaco04}.  When two DM halos merge, dynamical friction, tidal
stripping and tidal shocks on the satellite galaxies are followed
using the \citet{taffoni03} formulation. Similarly to S08, when a
satellite DM halo merges into a larger one, all of its hot gas is
shock heated according to the new halo potential and gets removed from
the satellite (thus implying that the corresponding satellite galaxy
does not receive any further cold gas supply).  Disc sizes are
computed using the model of \cite{mo98}, and bulge sizes are computed
assuming that kinetic energy is conserved in mergers.

A key ingredient in {\sc morgana} is the inclusion of a
self-consistent model for the accretion of gas onto supermassive black holes 
and the resulting AGN feedback modes (following \citealt{umemura01} and
\citealt{granato04}, see \citealt{fontanot06} for more details).  
This modeling assumes that the loss of angular momentum is the main
regulator of black hole accretion. This is triggered by the presence of gas 
in the bulge component, and the almost complete loss of angular momentum
of accreted gas is related to star formation activity.  Following
\citet{granato04}, star formation creates a reservoir of low angular
momentum gas which is then accreted at a rate regulated by the viscous
accretion time scale or by the Eddington limit. The nature of feedback
from the AGN depends on the accretion rate in Eddington units:
whenever this is higher than 0.01 (``bright mode''), the AGN can
trigger a massive galactic-scale wind \citep[see][]{monaco05} which
leads to the complete removal of the ISM from the galaxy, while at lower
accretion rates (``radio mode'')
the energy is ejected through jets that feed back on the hot halo gas
with an efficiency that scales with $V_c^2$, where $V_c$ is the halo
circular velocity. As a consequence, BH accretion requires some star
formation to be triggered, and feedback follows the onset of cooling
only after some time. The ejected energy heats the hot halo gas
component and quenches the cooling flow.
Galaxy SEDs, magnitudes and colours are obtained using the GRASIL
spectro-photometric code with radiative transfer for computing the 
effect of dust \citep{silva98}.

In Table 1, we present a brief description of each model.

\section{Results}
\label{sec:results}

In this section, we investigate the joint dependence of star formation
quenching on {\em stellar mass} and {\em halo mass}, in order to
try to contrain the physical mechanisms that are responsible for
quenching. We make use of two indicators of quenched star formation:
red optical colours, and low specific star formation rates ($SSFR
\equiv SFR/M_{\rm gal}$). Red optical colours are frequently used to isolate
``quenched'' galaxies; however, a red optical colour can arise from a
degenerate combination of an old stellar population, a high
metallicity, or strong dust extinction. SSFR based on UV-optical data
are a more sensitive probe of recent star formation.

In order to mimic the selection effects of the flux-limited
observational sample, we first assign redshifts to all the model
galaxies by placing an ``observer'' in a corner of the simulation
box.  We apply the same flux limit used in the observational catalogs
to the models by selecting only galaxies with apparent $r$-band
magnitude $r<17.77$. We then apply the $\rm{V_{max}}$ weighting
factor, just as we do with the galaxies in the observational
sample. Then, we exclude all halos that do not contain any galaxy
brighter than $^{0.1}M_r \ge -19.5 + 5\log h$, as these halos would
not be included in the group catalog.

We apply these selection criteria for all of the
model-data comparisons shown in the main text, unless otherwise noted.
We also present our main results for the models without these
selection criteria in the Appendix.

\subsection{Global Distribution Functions: colour and SSFR}
\label{sec:global}

In Fig.~\ref{fig:colourmag}, we present the global colour-magnitude
relations (CMRs) at $z=0.1$ for the observations and theoretical
models, along with the dividing line between the observed red sequence
and the blue cloud (sometimes called the ``green valley''). In this
figure, we show the results for the whole SDSS sample, regardless of
inclusion in the group catalog, and similarly we have not applied the
observational selection criteria to the theoretical models.
The magnitudes and colours are shown in the rest-frame $z=0.1$ system
defined by \citet{blanton03}. For the dL06 models, we used the
observed frame magnitudes at $z=0.1$, converted to absolute
magnitudes.  For the {\sc morgana} galaxies, we compute the absolute
$z=0.1$ magnitudes from the corresponding synthetic spectra.  For the
S08 models, when we computed standard $z=0.0$ frame colours and
magnitudes, we produced a good match to the observed colour-magnitude
relation expressed in the $z=0$ frame. However, when we computed the
$z=0.1$ system colours as described in \citet{blanton07}, we found it
necessary to apply a shift of 0.05 magnitudes to the $^{0.1}(g-r)$
colour to match the location of the observed $z=0.1$ system red
sequence. This is indicative of small differences in the shape of the
SED's in the semi-analytic models from the synthetic SED's used by
\citet{blanton03} for computing the k-corrections.  It should also be
noted that details in the population synthesis prescriptions, such as
chemical enrichment and dust extinction, may also cause noticeable
differences in colours (see Appendix for the effect of
dust). Therefore, reproducing the CMR quantitatively in semi-analytic
models is quite challenging.

The observations clearly show the familiar red sequence and blue
cloud, and these features are reproduced reasonably well in all of the
theoretical models, except the S08 no-AGN-feedback model. As has been
pointed out before \citep[e.g.,][]{croton06,cattaneo06}, semi-analytic
models without some kind of suppression of cooling in massive halos,
e.g. by AGN feedback, predict that massive galaxies are still
accreting plenty of cold gas at the present day, and therefore are
star-forming and blue, in conflict with observations. We do see subtle
differences in the structure of the CM distribution for the different
models, for example, the S08-halo-quenching model produces very few
low-luminosity red galaxies, and the dL06 model produces a very strong
bimodality. Also, we notice that the red sequence in the
halo-quenching model is slightly bluer compared with the S08 fiducial
model.  This is because the central galaxies in the halo-quenching
model stop forming stars once their halo becomes more massive as
$10^{12} h^{-1}M_\odot$. In general, this happens at an earlier epoch
than the ``radio mode'' feedback is able to quench star
formation. Because chemical evolution is also halted when star
formation is quenched, the massive galaxies in the halo-quenching
model do not become as enriched as do the galaxies in the fiducial
model (see S08). We could have adjusted for this by increasing the
chemical yield, but we chose to leave all the free parameters fixed in
both models to allow a direct comparison.

The behavior of {\sc morgana} is somewhat different from that of the
other models. A clear red sequence is present at intermediate and
faint magnitudes, but it fades away at bright magnitudes, in contrast
to the observations. Luminous galaxies tend to be blue. We interpret
this result as an inefficient quenching of star formation through
``radio-mode'' feedback in {\sc morgana}. In fact, as explained in
\S.~\ref{sec:models}, in this model the AGN heating switches on only
after some cooled gas has already started forming stars in the host
galaxy. Obviously the residual activity is stronger for longer time
delays between the onset of cooling flows and the accretion onto the
central black hole.  On the other hand, the blue cloud seems depleted. As
shown in Fontanot et al. (2008, in prep), the depletion of the blue
cloud is connected in part to the ``satellite overquenching problem'',
discussed further in \S5, caused by the too-efficient strangulation of
satellite galaxies. However, we shall see later that {\sc morgana}
also produces too few low mass blue central galaxies. This is likely
due to strong supernova feedback.

\begin{figure*}
\begin{center}
\includegraphics[width=8cm]{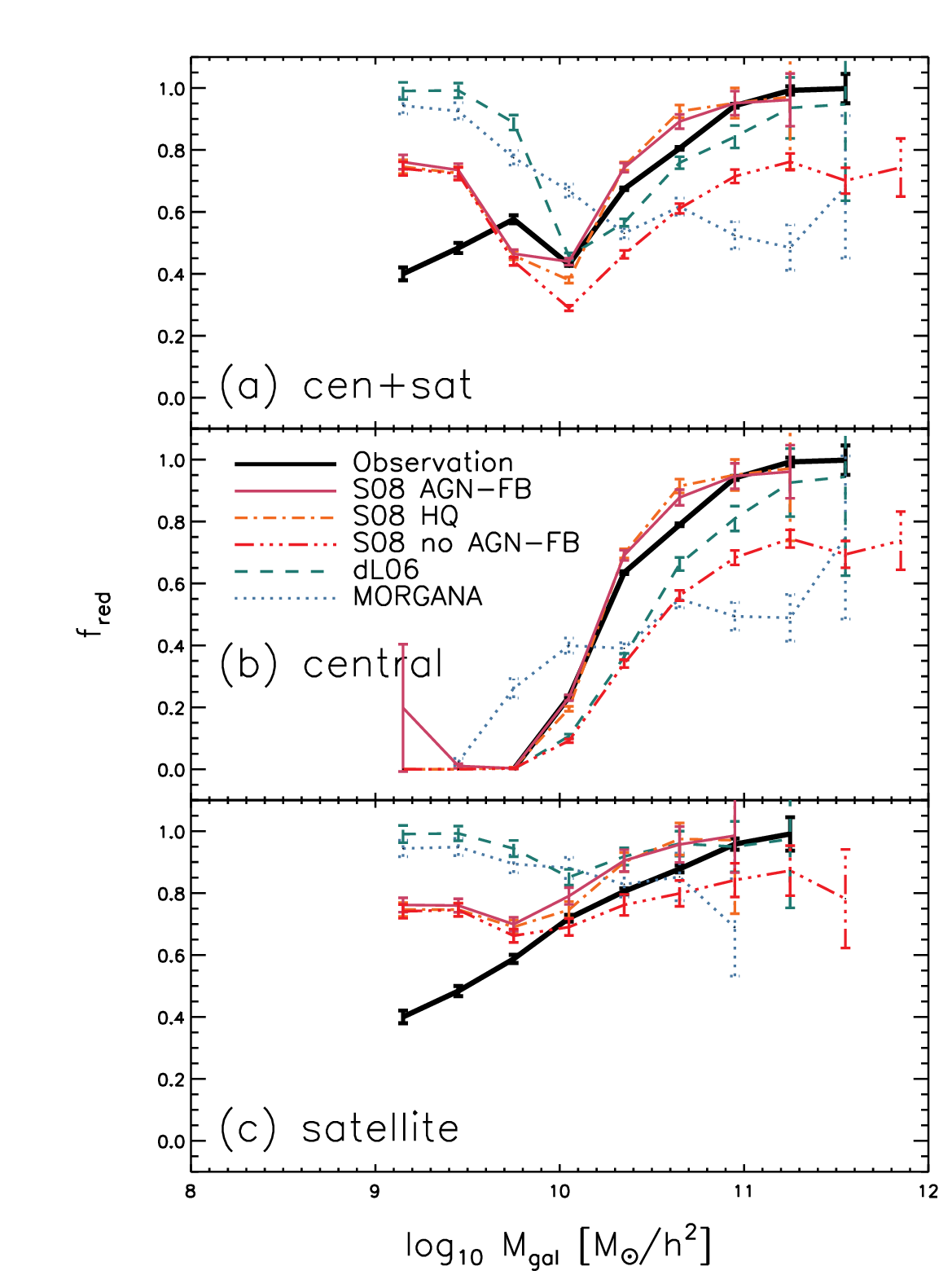}
\includegraphics[width=8cm]{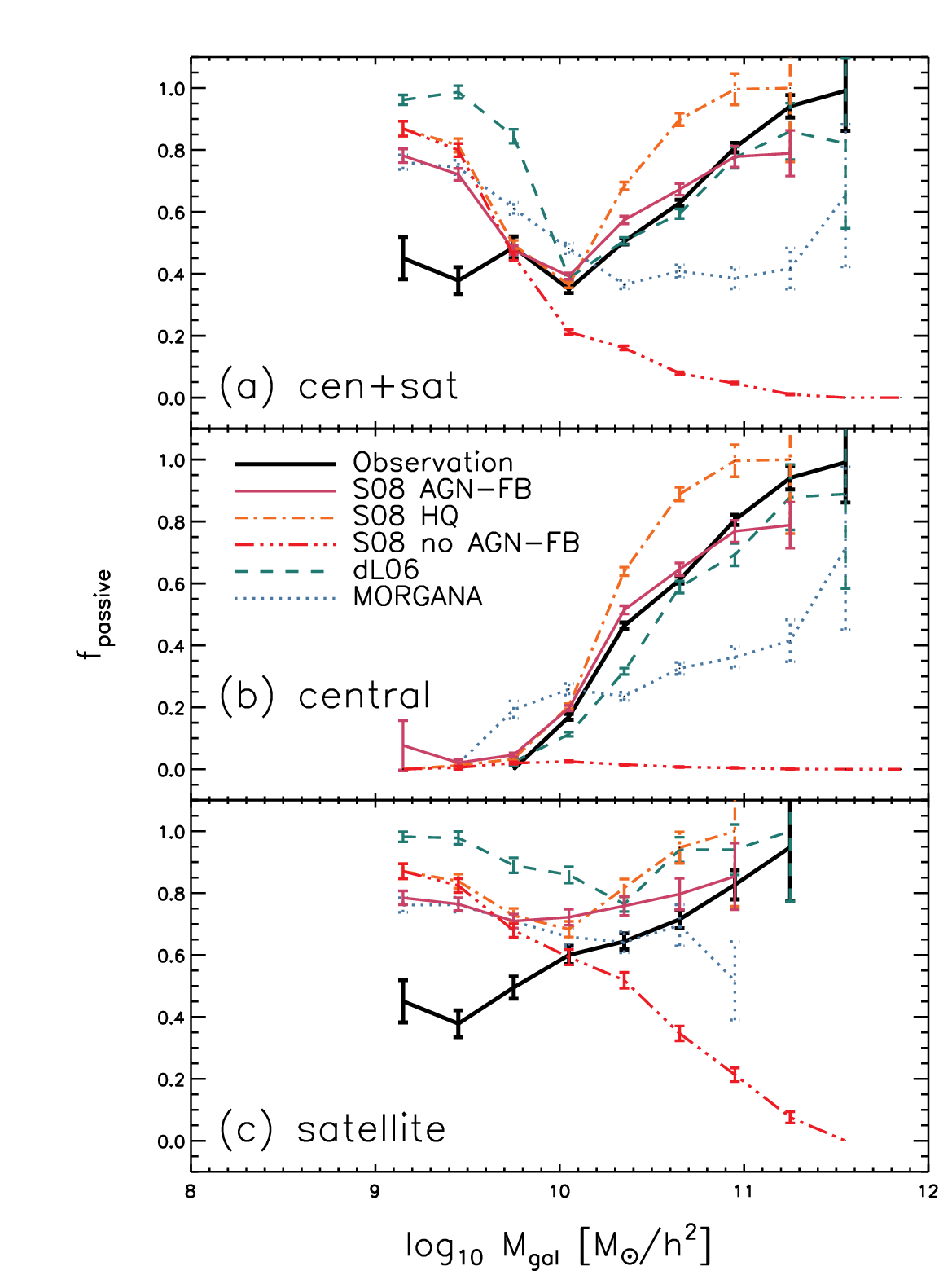}
\caption{The fraction of red galaxies ($\rm{f_{red}}$; left) and
  passive galaxies ($\rm{f_{passive}}$; right) as a function of galaxy
  stellar mass.  The empirical data are shown by a black solid
  line, and coloured lines show the results of the theoretical models,
  as indicated on the plot.  We show the dependence for all galaxies
  (top), and for central (middle), and satellite (bottom) galaxies
  seperately. Each point contains at least 10 galaxies.  Most of the
  SAMs reproduce the trend for central galaxies reasonably well, but
  predict a much larger fraction of small-mass red/passive satellite
  galaxies than are observed. The error bars indicate Poisson errors.
}
\label{fig:fred_mgal}
\end{center}
\end{figure*}

\begin{figure*}
\begin{center}
\includegraphics[width=8cm]{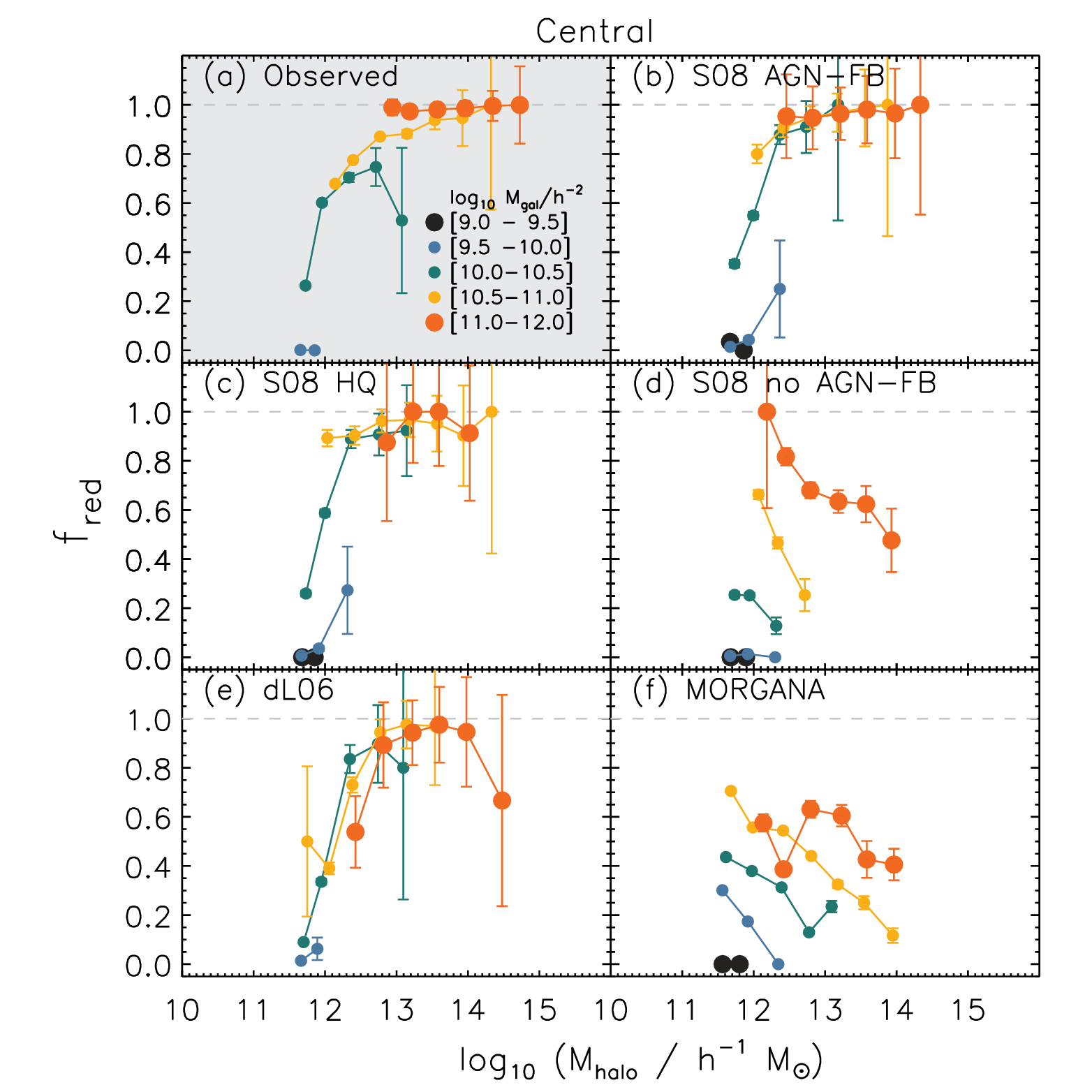}
\includegraphics[width=8cm]{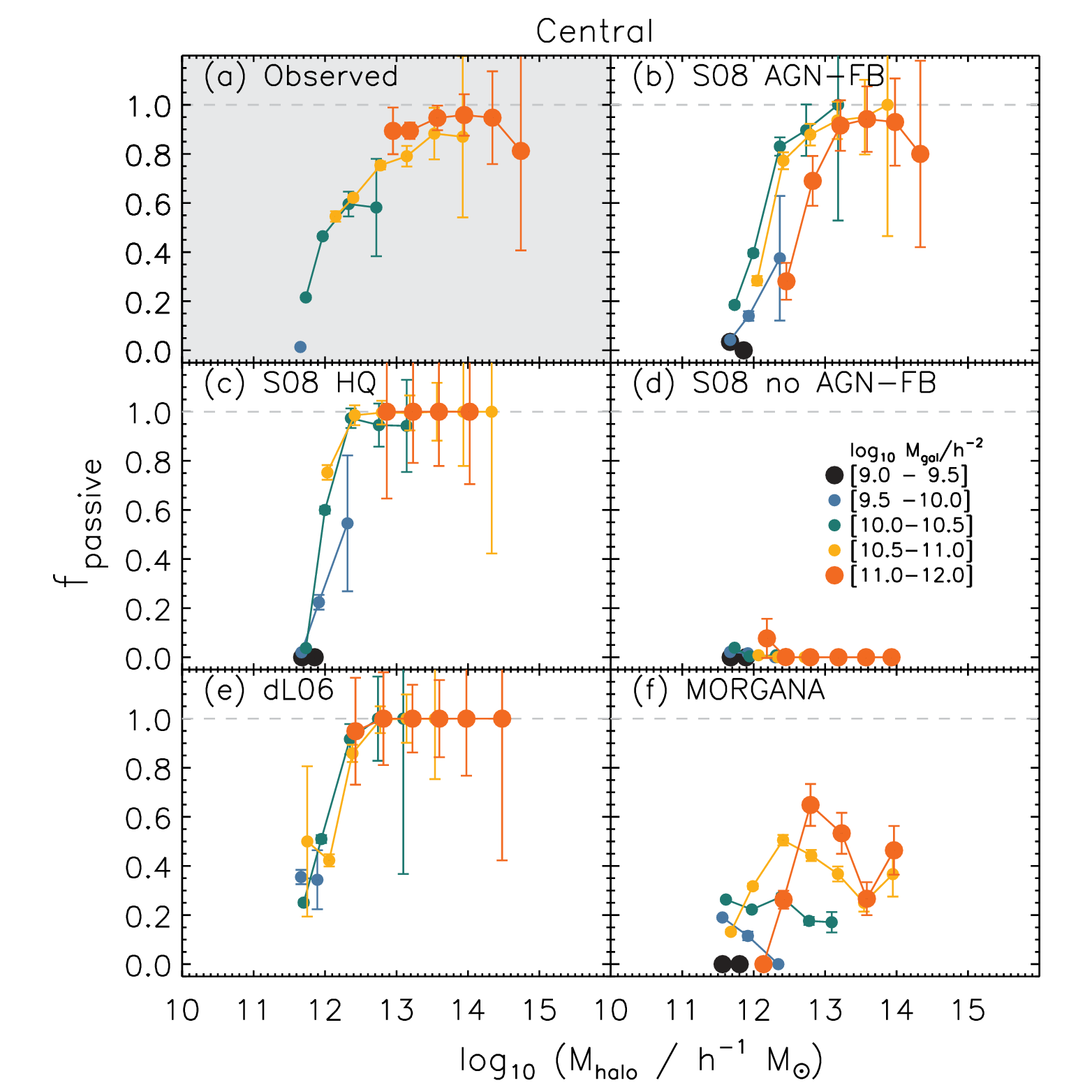}
\includegraphics[width=8cm]{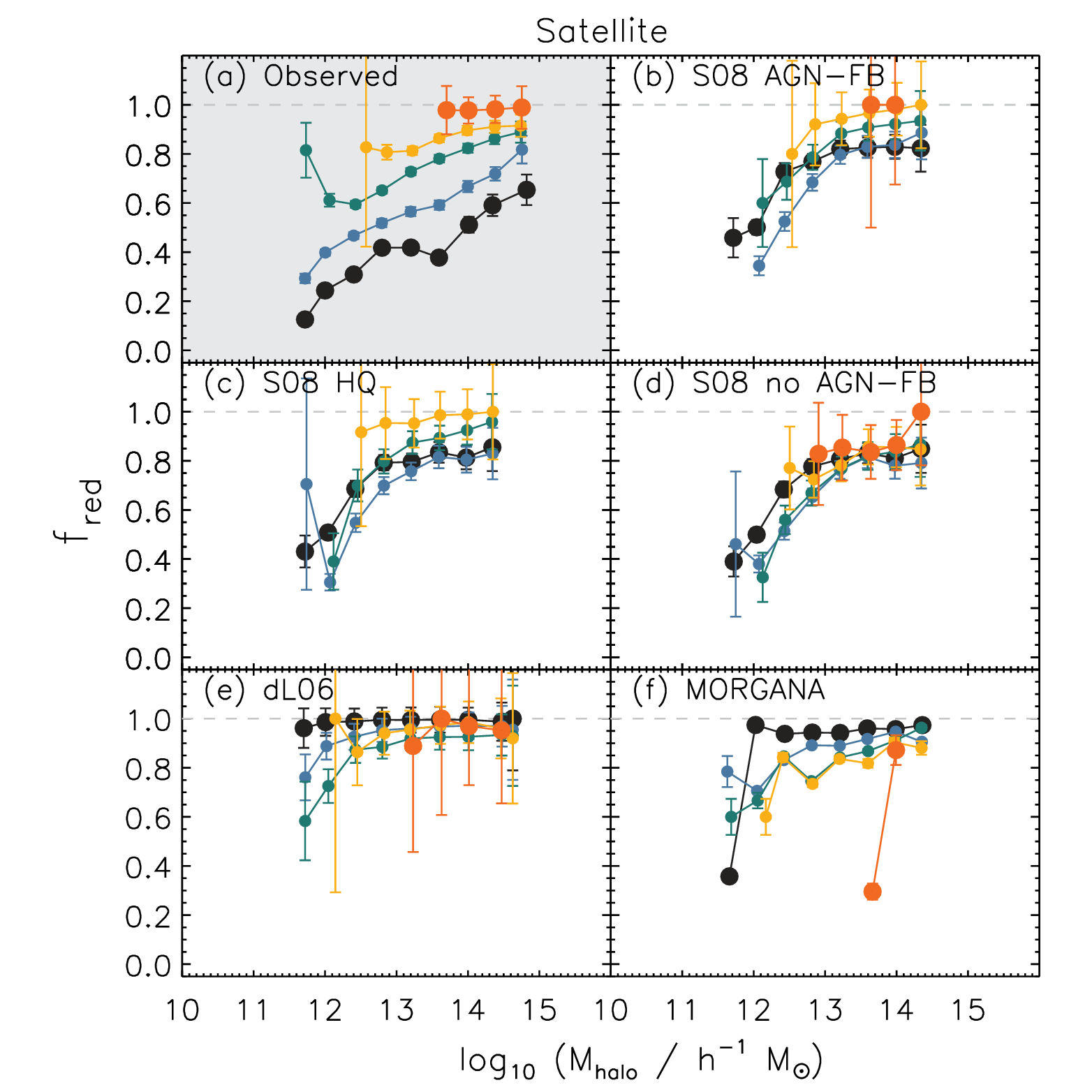}
\includegraphics[width=8cm]{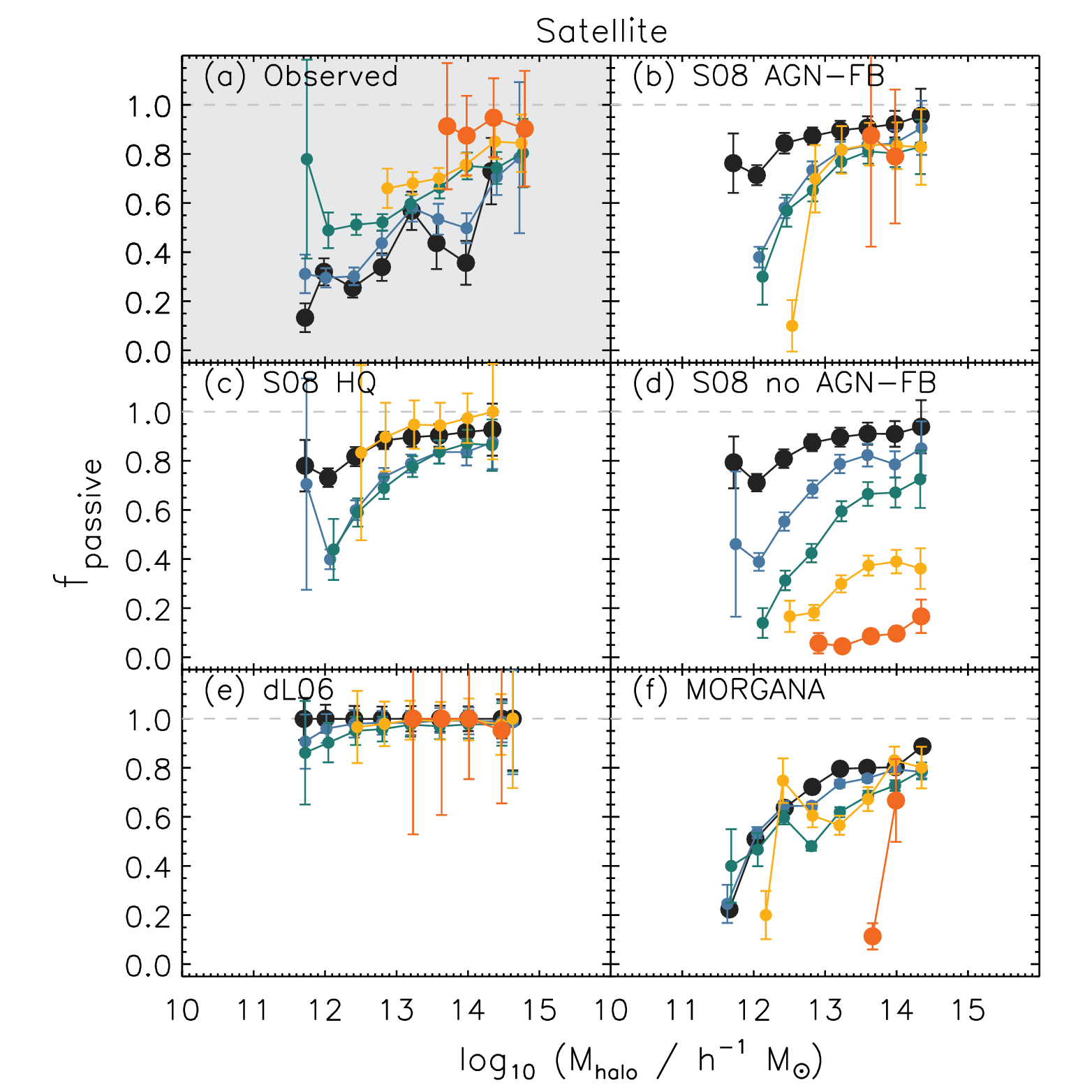}
\caption{The red fraction $\rm{f_{\rm red}}$ (left) and passive
  fraction $\rm{f_{passive}}$ (right) as a function of halo mass, for
  different stellar mass bins, as indicated by different colours and
  symbol sizes (see plot legend). In each plot, results are shown for
  the observed SDSS or GALEX+SDSS group catalog (top left) and the
  five models as indicated on the plot. We present the results for
  central ({\it top set of panels}) and satellite ({\it bottom set of
    panels}) galaxies seperately.  Each point contains at least 5
  galaxies. The SAMs reproduce the main trends reasonably well for
  central galaxies, but satellite galaxies do not show the correct
  trend of $f_{\rm red}$ with stellar mass. }
\label{fig:fredmgalmhalo1}
\end{center}
\end{figure*}

\begin{figure*}
\begin{center}
\includegraphics[width=8cm]{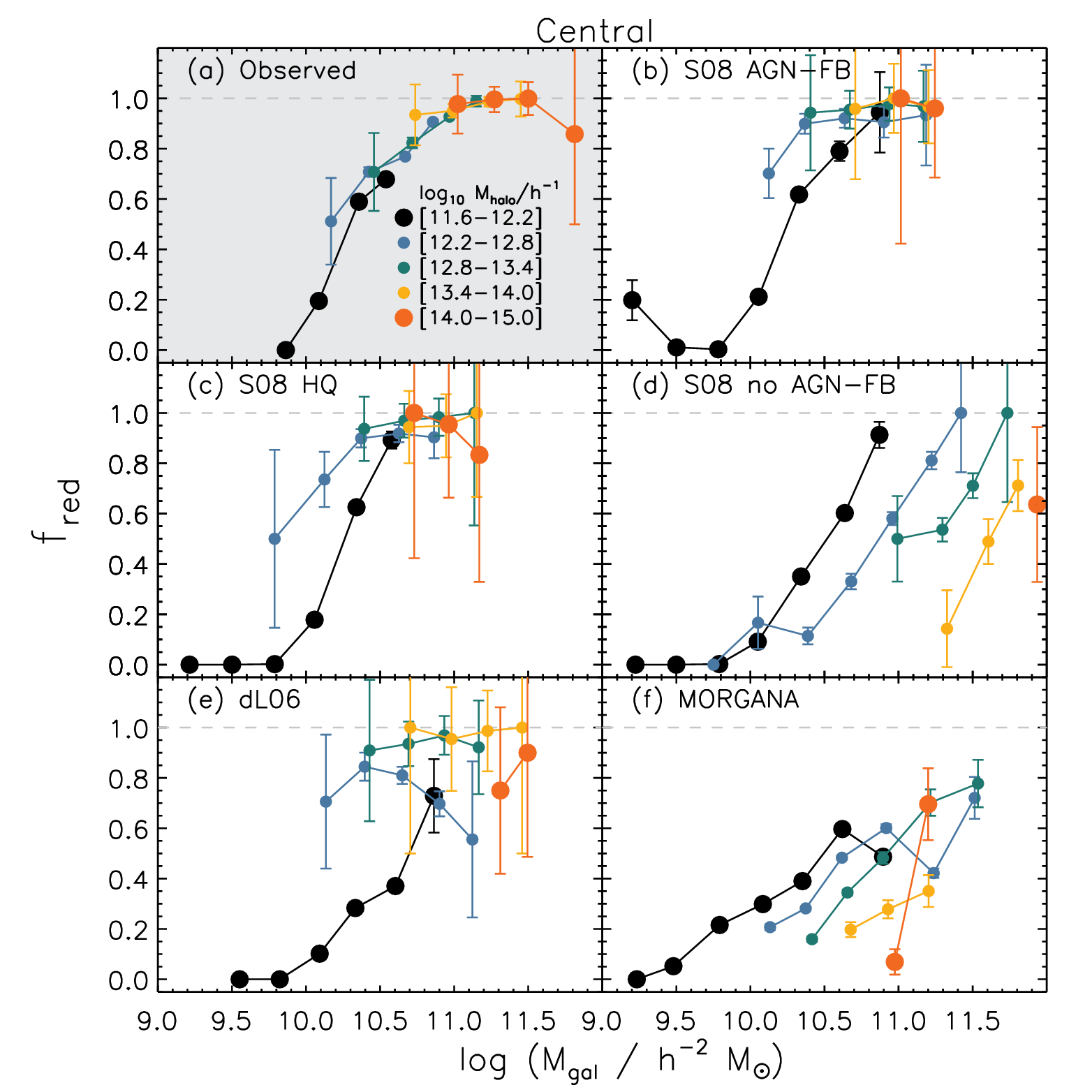}
\includegraphics[width=8cm]{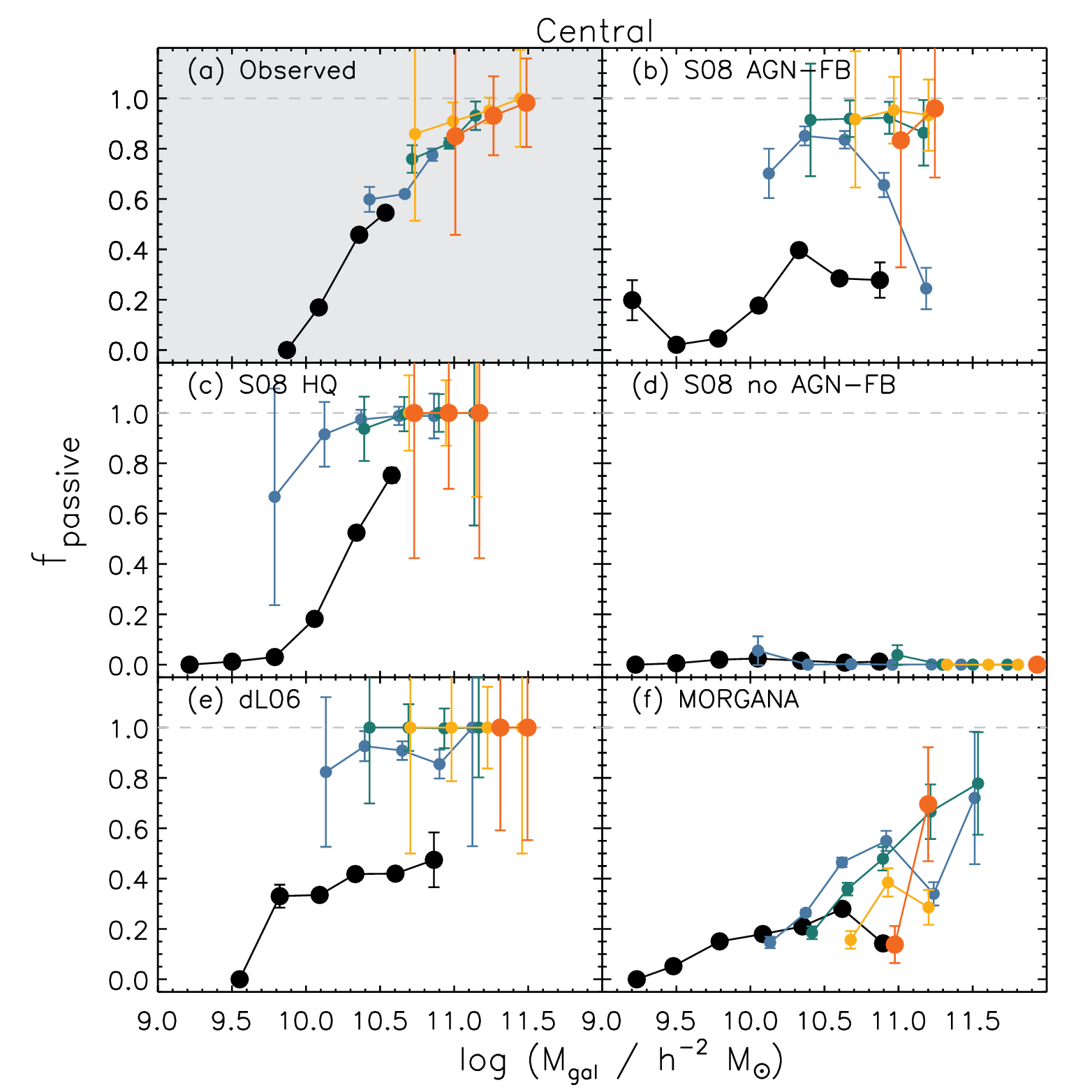}
\includegraphics[width=8cm]{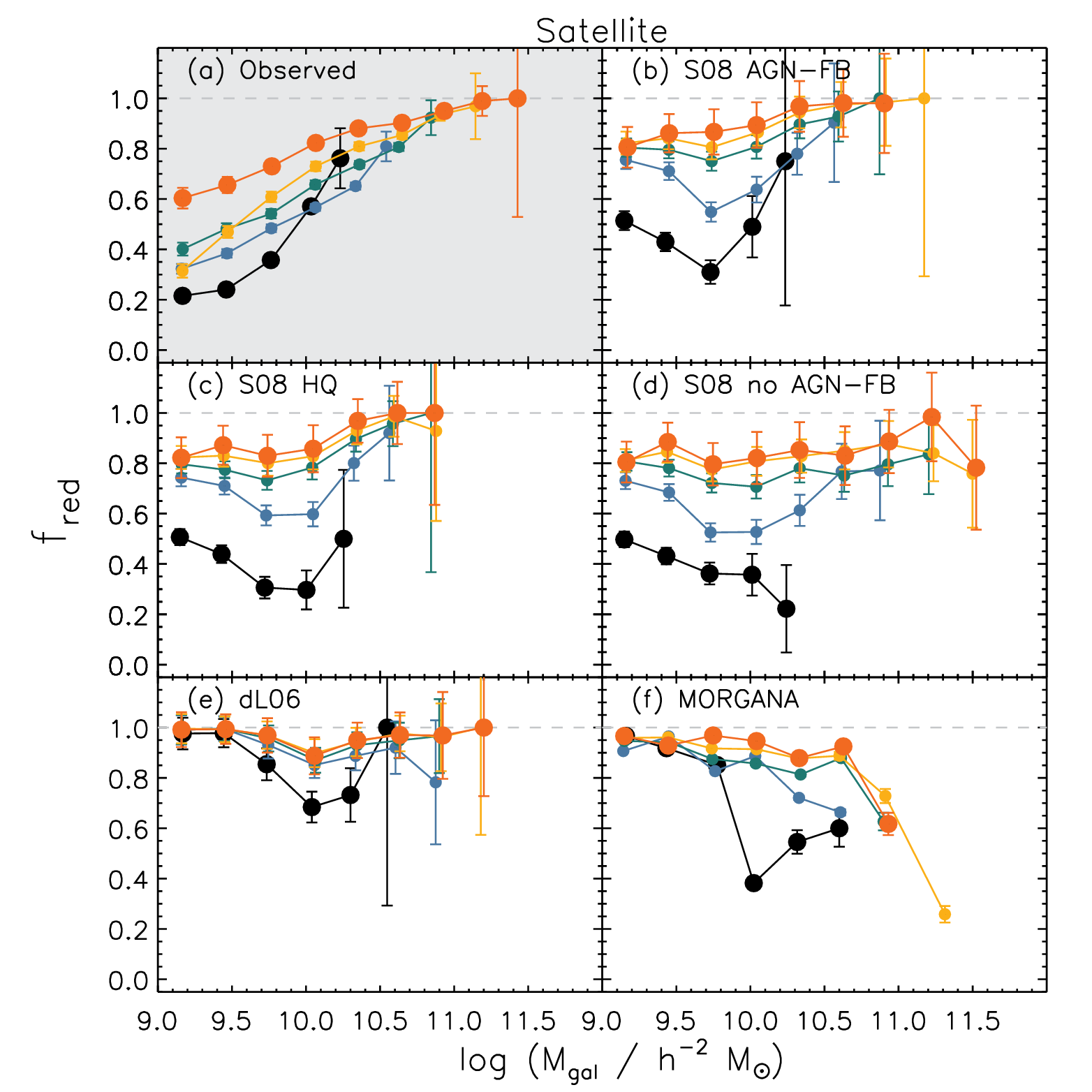}
\includegraphics[width=8cm]{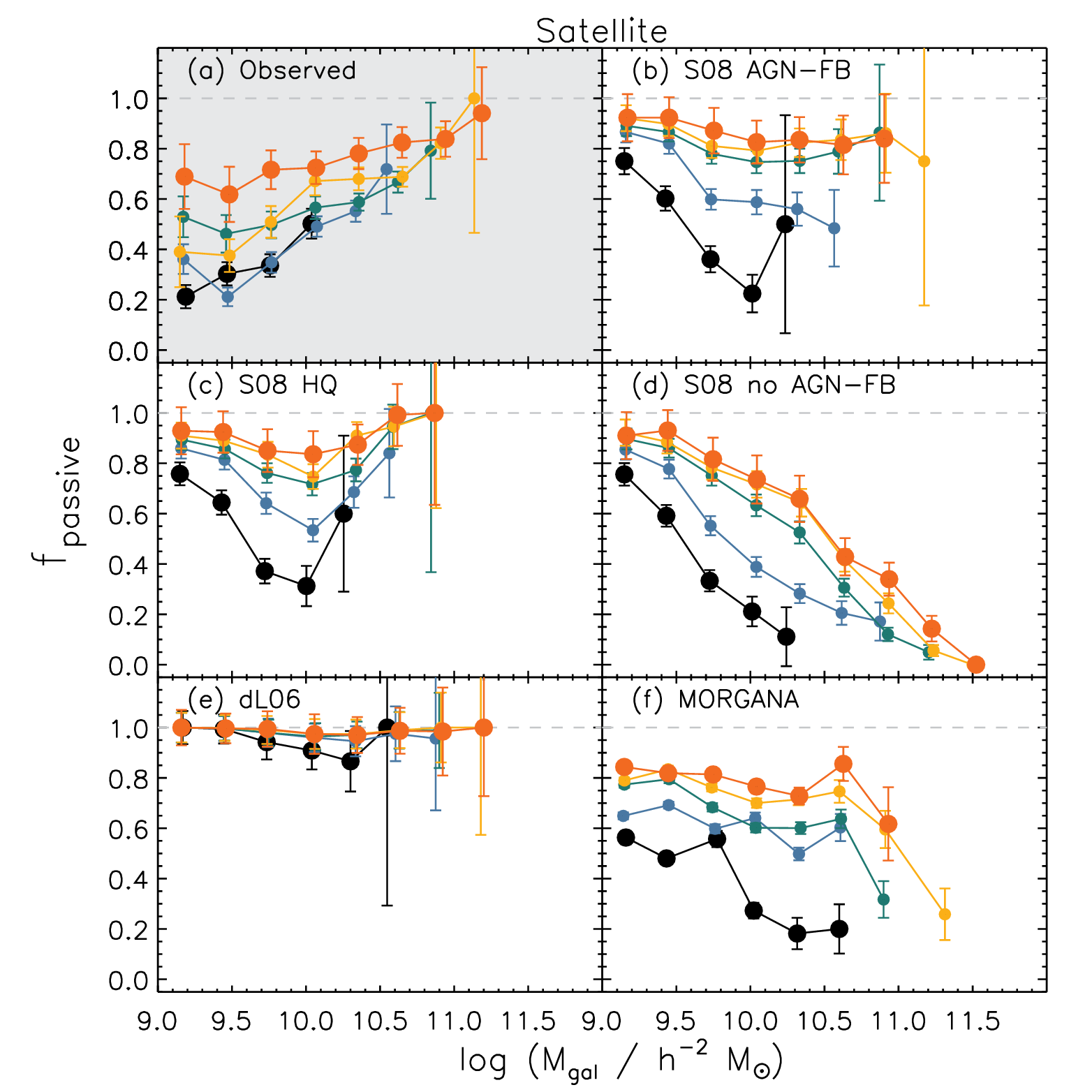}
\caption{The red fraction $\rm{f_{\rm red}}$ (left) and passive
  fraction $\rm{f_{passive}}$ (right) as a function of stellar mass,
  for different halo mass bins, as indicated by different colours and
  symbol sizes (see plot legend). The rest of the plot details are as
  in Fig.~\protect\ref{fig:fredmgalmhalo1}.  }
\label{fig:fredmgalmhalo2}
\end{center}
\end{figure*}

Given the strongly bimodal colour distribution exhibited by both the
empirical data and most of the models, it is natural to define a
dividing line between the two populations, and to investigate the
fraction of galaxies in the red population, \fred, as
representing quenched objects. We adopt the same demarcation line
between red and blue galaxies as Weinmann et al.  (2006a, hereafter
W06a):
\begin{equation}
^{0.1}(g-r) = 0.7 - 0.032 ( ^{0.1} M_r - 5 \log_{10} h + 16.5)
\label{def_colour}
\end{equation} 
as shown in Fig.~\ref{fig:colourmag}.

We can also define the specific star formation rate (SSFR), as the
present star formation rate divided by the stellar mass of the galaxy
($=SFR/M_{\rm gal}$), and plot a similar diagram in terms of SSFR and
mass. In Fig. \ref{fig:ssfr}, we show the conditional probability
distribution for SSFR as a function of stellar mass, as derived from
the GALEX+SDSS observations, and for the models. The star formation 
rate is averaged
over the past 100 Myr. Note that the number of galaxies in the sample
is roughly 8 times smaller than the sample shown in the
colour-magnitude distribution plot, because the field coverage of GALEX
survey used in the Salim et al.  study is not as wide as that of the
SDSS. In a similar manner, we define a cut in SSFR to separate
``active'' star forming galaxies from ``passive'' ones. We define a
galaxy as ``passive'' if the following condition is met:
\begin{equation}
\log_{10} SSFR \le  -9 - 0.2 \log [\mgal/h^{-2}M_\odot]
\label{def_ssfr}
\end{equation}

Note here that our criterion is based on galaxy stellar mass in order
to make a more direct comparison with the theoretical models. It is
also worth noting that our criterion roughly corresponds to $SFR \sim
1 M_\odot \rm yr^{-1}$ at \mgal$\sim10^{11}h^{-2}\msun$ and $SFR
\sim0.1 M_\odot \rm yr^{-1}$ at \mgal$\sim10^{9.5}h^{-2}\msun$.  Below
0.1$M_\odot \rm yr^{-1}$, SFR obtained from multiband photometry may
not be robust due to the degeneracy of burst time and the mass
fraction of a young population \citep[e.g.][]{kaviraj07}.

Our demarcation is comparable to that of W06a when their criterion,
which is based on the $^{0.1}r$-band magnitude, is lowered by 0.6 dex.
However, SFR estimates from emission lines, which were used by W06a,
trace only very recent star formation, whereas the UV used here traces
star formation over a longer timescale ($\sim$ 1 Gyr). Therefore our ``passive''
sample is not directly comparable to that of W06a, though the results
are qualitatively similar.
The amount of UV flux in the passive galaxies is very small and can
still be consistent with the amount of UV flux that can be
produced by old stars such as low-mass horizontal-branch stars (see 
\citealt{yi05} for details).  In this regard it is justifiable to call
them ``passive''.  

In Fig.~\ref{fig:ssfr}, we see a well-defined ``star-forming sequence''
that produces the blue sequence in the traditional CM diagram, but the
``quenched'' population that produces the tight red sequence is quite
spread out in SSFR. This is simply a reflection of the relative
insensitivity of optical colours to small amounts of recent star
formation. The precise values of SSFR at low levels of star formation cannot be
estimated very accurately from the data, and therefore one should not
take too seriously the position of the quenched galaxies in the SSFR
vs. \mgal\ plot. Again, the models with AGN feedback (or halo
quenching) qualitatively reproduce the empirical distribution
reasonably well, but show some interesting differences. The dL06 model
has a strongly-bimodal distribution of SSFR, with most galaxies living
either on the star-forming sequence or being completely quenched. In
contrast, the S08 model has a larger population of ``semi-quenched''
galaxies at intermediate values of SSFR. The {\sc morgana} model
produces an even broader distribution of SSFR, with many quenched
low-mass galaxies.

\subsection{Dependence of Star Formation Quenching on Stellar Mass}

Fig.~\ref{fig:fred_mgal} shows the dependence of \fred\ and \fpass\ on
galaxy stellar mass.  We show \fred\ and \fpass\ vs. stellar mass
for all galaxies (upper panels), and for central and satellite
galaxies separately (middle and lower panels, respectively). Although
we do not specifically use the information from the group catalog in
the uppermost panel, in all panels we use only the SDSS galaxies that
are included in the group catalog, and apply the group catalog-like
selection criteria to the theoretical models, as described at the
beginning of this section.

In the empirical data, \fred\ shows a strong dependence on galaxy mass
for both centrals and satellites, in the well-known sense that
low-mass galaxies are largely blue, whereas massive galaxies are more
likely to be red. It appears that \fred\ is a steeper function of
stellar mass for central galaxies than for satellites.
In general, the red galaxy fraction is higher for satellites for a
fixed galaxy mass. Similarly, \citet{vdb08a,vdb08b} found
that satellite galaxies have redder mean colours than centrals at a
fixed stellar mass. The trends appear qualitatively very similar when
we consider \fpass\ as a function of stellar mass.

The S08 fiducial model, S08 halo-quenching model, and dL06 model all
reproduce the trends in \fred\ and \fpass\ with stellar mass quite
well for {\em central} galaxies. Note the similarity of the
predictions of the S08 fiducial and halo-quenching model. The S08 no
AGN feedback model, as we have already seen, predicts the presence of too
many massive blue central galaxies.  
{\sc morgana} reproduces the sense of the trend of the red fraction
with stellar mass, but slightly overproduces red galaxies at low
stellar masses, and significantly overproduces blue galaxies at high
stellar masses.

All of the models badly overproduce the number of low mass red
satellites, and predict too weak a trend of \fred\ with stellar mass
for these objects. The dL06 models predict a nearly flat run of
\fred\ with stellar mass for satellites, with values that are much too
high (close to unity) compared with the empirical values.  At the
highest satellite masses, there is {\em too high} a fraction of blue
galaxies in the {\sc morgana} model and the S08 no-AGN-feedback model.
We can see by comparing the S08 no-AGN-feedback model with the
fiducial model that AGN feedback does not affect galaxy colours below
stellar masses of $\sim 10^{10} h^{-2}\msun$; therefore {\em the
excess of low-mass red satellites is not connected with AGN
feedback.}

The conclusions we would draw from the comparison with \fpass\ are
qualitatively similar, though quantitatively somewhat different. For
example, the dL06 model shows a better quantitative match to the
\fpass\ data than to \fred\ for central galaxies. The S08 fiducial and
halo-quenching models produced almost indistinguishable results for
\fred(\mgal), but significantly different results for
\fpass(\mgal). This is due to the age-metallicity degeneracy --- as we
discussed in \S\ref{sec:global}, massive galaxies in the halo-quenching
model are more metal poor than in the fiducial model. Therefore,
although they are older (as seen in the \fpass\ diagram), their
optical colours are similar. Similarly, the large difference between
\fred\ and \fpass\ for the S08 no AGN-feedback model is due to dust
extinction: in this model, massive galaxies are actively star forming,
and therefore extremely dusty (see Appendix for more discussion of the
effects of dust extinction). 

The somewhat lower fraction of red/passive satellite galaxies in the
S08 models is due to the inclusion of tidal destruction, which is not
included in the dL06 or {\sc morgana} models. In the S08 model,
satellites that orbit within their host halo for a long time can
eventually become tidally destroyed, and their stars are added to a
``Diffuse Stellar Halo'' (see S08). Naturally, in the absence of tidal
destruction, these satellites exhaust all of their gas and become very
red. However, we see here that although the inclusion of tidal
destruction helps reduce the excess of low-mass red satellites in the
models, a significant discrepancy still remains.

One might then wonder whether increasing the efficiency of tidal
destruction could completely solve the satellite 'over-quenching'
problem that we see here. We do not believe that this is a viable
solution, for several reasons. First, the tidal disruption model
used by S08 was tuned to match the sub-structure mass function for
very high-resolution N-body simulations. S08 also showed that their
models correctly reproduce the total number of low-mass galaxies
(i.e. the faint end slope of the stellar mass function), although
the model overproduces low-mass bulge-dominated (red/passive)
galaxies and underproduces low-mass disc-dominated (blue/active)
galaxies. Significantly increasing the efficiency of tidal
destruction would be in conflict with the N-body results and would
also produce an overall deficit of low-mass galaxies. Put another
way, tidal destruction can remove low-mass red satellites but cannot
increase the number of low-mass blue satellites.

\subsection{Dependence on Stellar Mass and Halo Mass}

In this section we explore the dependence of \fred\ and \fpass\ on DM
halo mass and stellar mass.  In Fig.~\ref{fig:fredmgalmhalo1} we show
\fred\ and \fpass\ as a function of halo mass, for different bins in
stellar mass. We show the results for \fred\ and \fpass\ for central
galaxies (top row), and for satellite galaxies (bottom row). In
Fig.~\ref{fig:fredmgalmhalo2}, we show a similar plot, but this time
with the galaxy stellar mass plotted on the x-axis, and different bins
in halo mass shown by the different colors.

In the empirical data, we see that for massive $\mgal > 10^{11}
h^{-2}\msun$ {\em central} galaxies, there is no significant
dependence of \fred\ on halo mass (environment) for fixed stellar
mass. For intermediate mass galaxies ($10^{10} h^{-2} < \mgal <10^{11}
h^{-2}\msun$), the dependence on halo mass appears to be stronger, but
there is a limited region of overlap in galaxies with different
stellar masses that occupy halos of the same mass. This is because
there is a fairly tight correlation between halo mass and stellar
mass. Similar results are again obtained for \fpass.

The S08 fiducial and dL06 models, both of which include AGN feedback,
do reasonably well at reproducing the overall trends for central
galaxies, as does the S08 halo-quenching model. We see a hint of a
dependence on stellar mass in intermediate mass halos
($10^{11.5}h^{-1}\msun \lesssim \mhalo \lesssim 10^{12.5}
h^{-1}\msun$) in the S08 models, while in the dL06 model, the
dependence seems to be almost solely on halo mass. However,
interestingly, we see almost the same stellar mass dependence in the
S08 fiducial and halo-quenching models, while we know that the
quenching mechanism is a pure function of halo mass in the
halo-quenching model.  The S08 no-AGN feedback model shows the correct
trend with stellar mass at fixed halo mass (more massive galaxies have
higher \fred), but the opposite trend with halo mass (more massive
halos have lower \fred).  Interestingly, {\sc morgana} shows similar
trends to the S08 no-AGN-feedback model: this implies that the current
implementation of AGN feedback in this model is insufficient to fully
cure the ``star formation quenching'' problem.

Considering \fpass, we see somewhat different behavior.  The S08 no
AGN-feedback model produces no quenched galaxies at all in terms of
\fpass. The S08 and dL06 models appear surprisingly similar, both
showing almost a pure halo mass dependence (no significant dependence
on stellar mass).  On the other hand, {\sc morgana} predicts
a significant dependence of \fpass\ on stellar mass, while showing
overall low values of \fpass.

Observed satellite galaxies show an \fred\ dependence that is very
similar to that of centrals for massive galaxies ($\mgal \gtrsim
10^{10.5} h^{-2}\msun$). For intermediate and low mass satellites,
\fred\ shows some dependence on both galaxy stellar mass and halo
mass, but does not show the very sharp drop over intermediate halo
masses ($11.5 \lesssim \mhalo \lesssim 12.5 h^{-1} \msun$) seen in the
central population. In the semi-analytic models, we see that
\fred\ for the satellites does not have a strong enough dependence on
\mgal\ at fixed halo mass. 

The stellar mass dependence of \fpass\ for satellites in the empirical
sample is not as clear as it was in terms of \fred. This may be in
part due to the smaller size of the GALEX-SDSS matched sample used to
obtain \fpass. The S08 models now all show a clear {\em inverted}
trend: \fpass\ is higher for lower mass galaxies (the opposite of the
empirical trend). For the dL06 model, nearly all satellites are
passive regardless of their stellar mass or halo mass.  In {\sc
  morgana}, satellite properties are a weak function of stellar mass,
and the values of \fpass\ are overall too high.

Based on Fig.~\ref{fig:fredmgalmhalo1} alone, we might be tempted to
conclude that quenching is primarily a function of halo mass. However,
Fig.~\ref{fig:fredmgalmhalo2} shows that quenching could equally well
be considered to be primarily a function of stellar mass. We conclude
that, especially for central galaxies, the degeneracy between stellar
mass and halo mass is too strong to reach a firm conclusion on this
point.

\begin{figure*}
\begin{center}
\includegraphics[width=8cm]{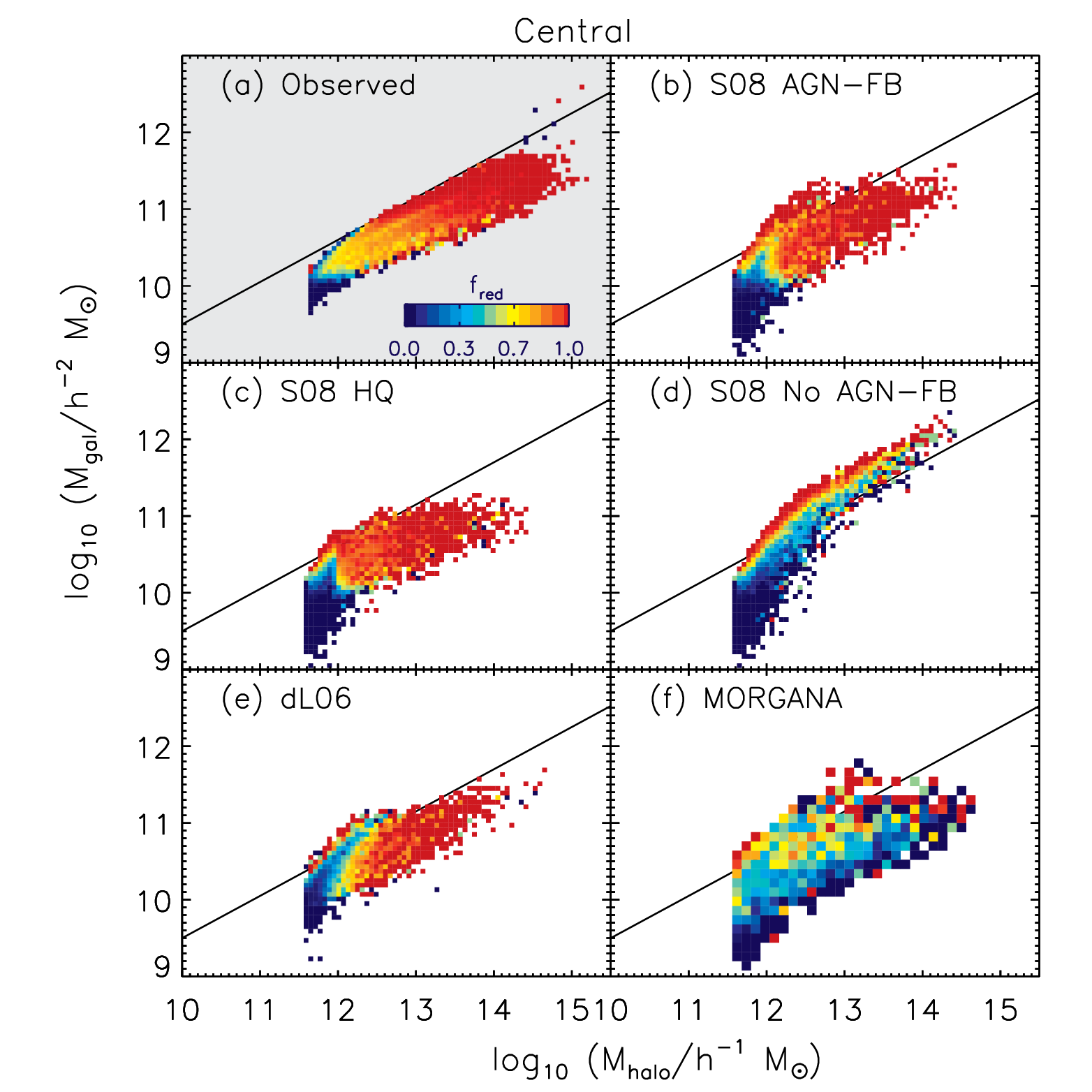}
\includegraphics[width=8cm]{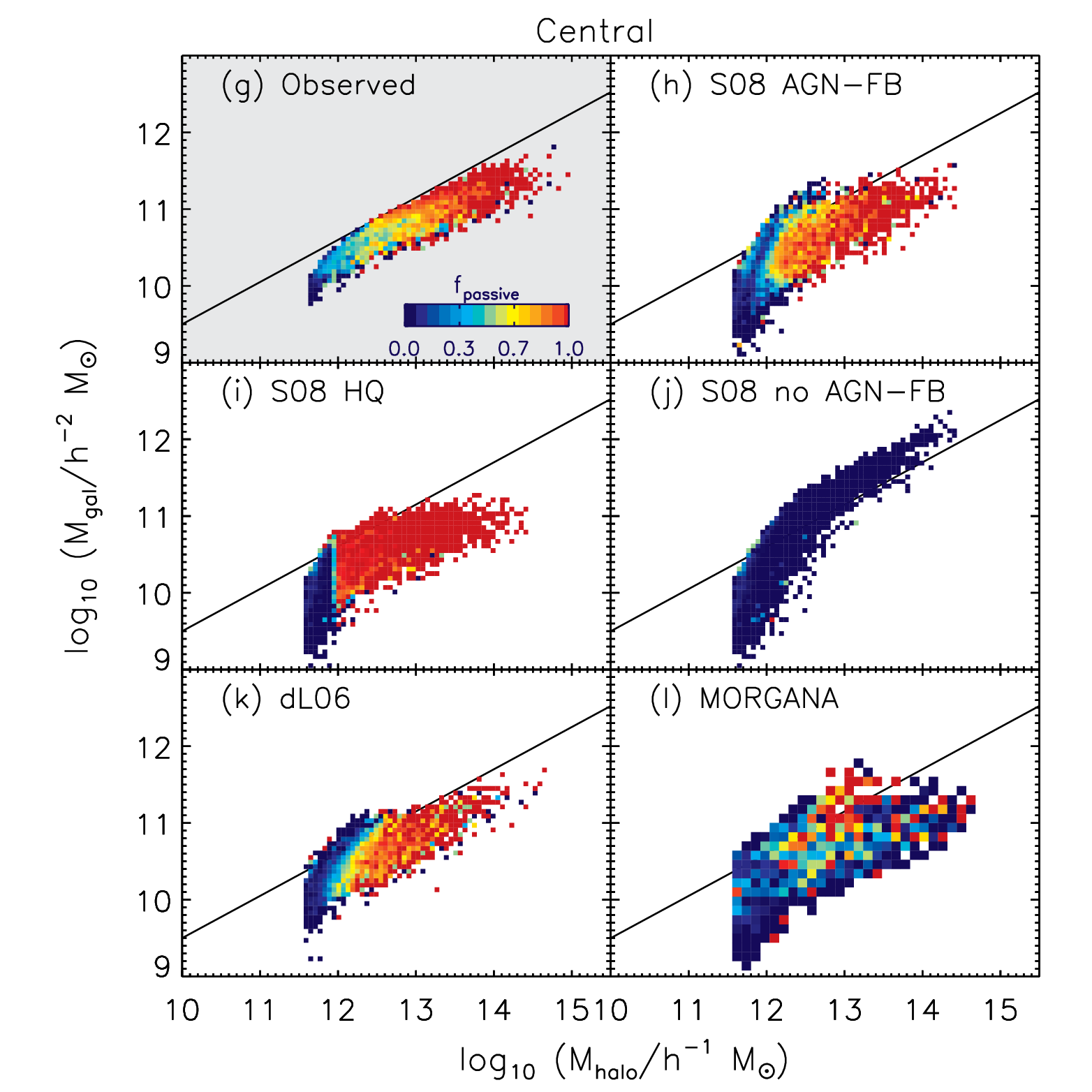}
\includegraphics[width=8cm]{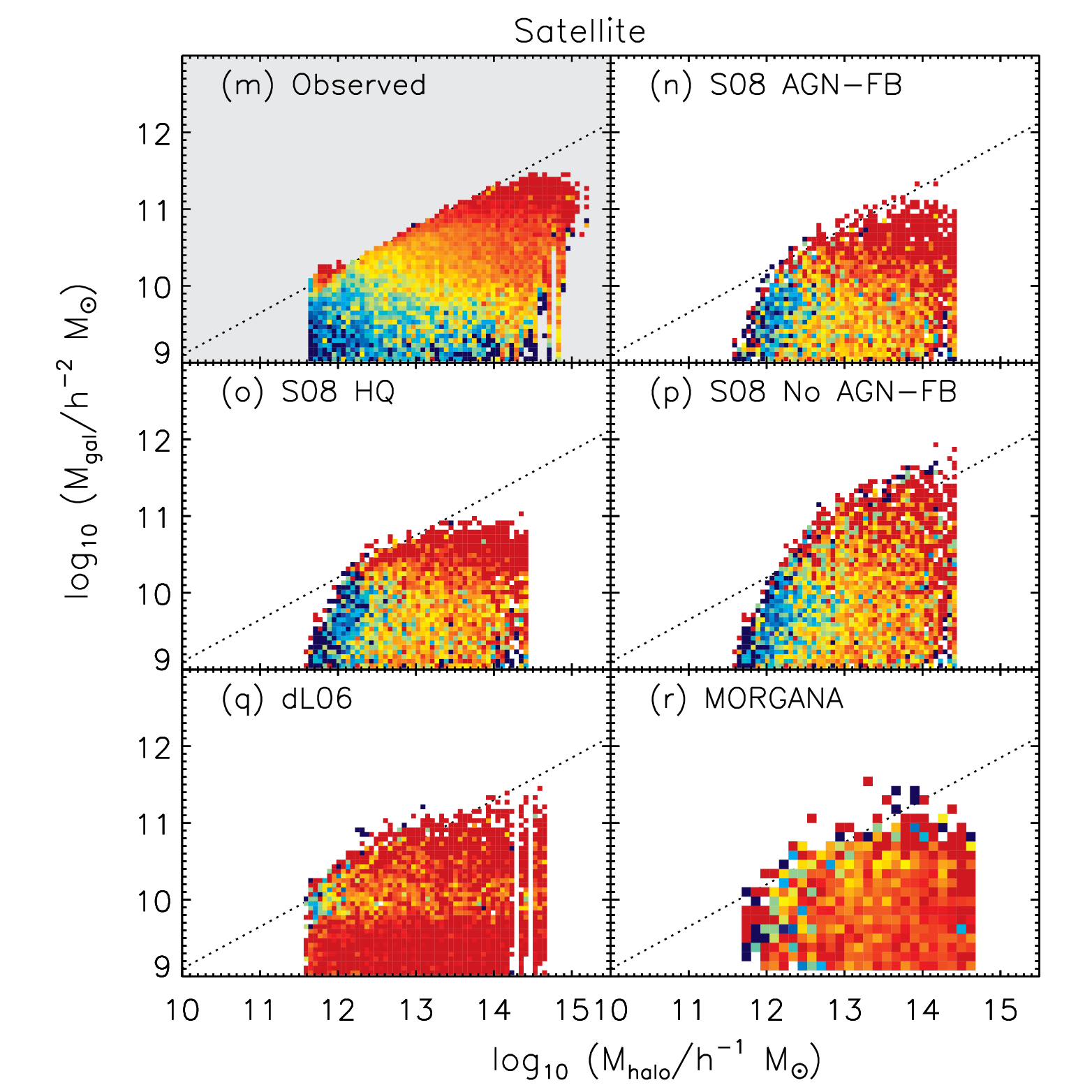}
\includegraphics[width=8cm]{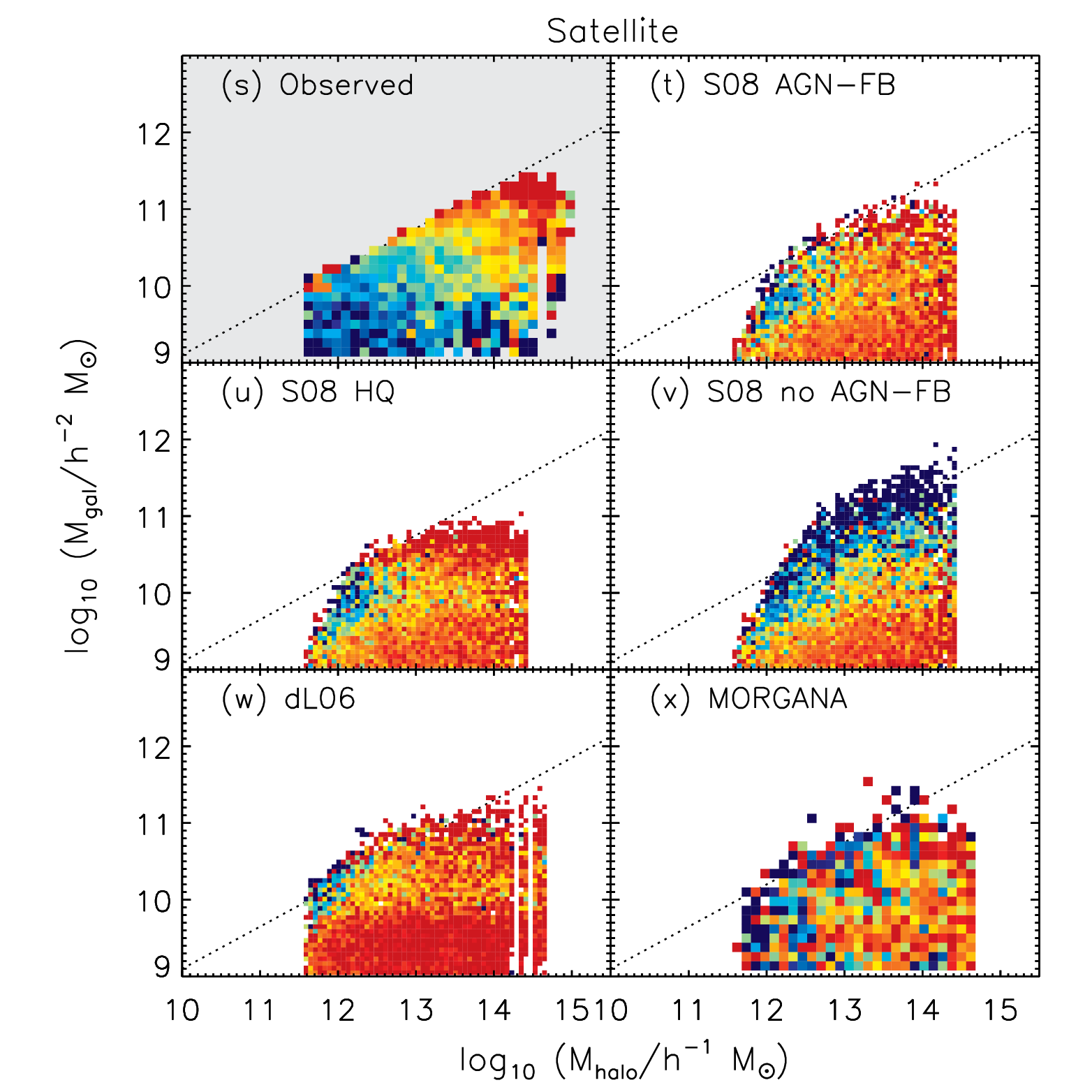}
\caption{The fraction of ``red'' galaxies (\fred; left) and passive
  galaxies (\fpass; right) for central ({\it top set of plots}) and
  satellite ({\it bottom set of plots}) galaxies in the (\mhalo,
  \mgal) plane. The fraction of red/passive galaxies in a given pixel
  in (\mhalo, \mgal) is indicated by the colour, where red colours
  indicate a higher red/passive fraction, as shown in the scale. To
  guide the eye, we draw a solid (dotted) line showing the approximate
  upper envelope of the central (satellite) galaxy mass distribution
  for the observational group catalog, and repeat this same line on
  every panel. Central and satellite galaxies in the observational
  group catalogs show noticably different joint dependencies on
  stellar mass and halo mass. The models with AGN feedback
  qualitatively reproduce the trends for central galaxies, but do not
  reproduce the empirical trends for satellites. }
\label{fig:fred2d}
\end{center}
\end{figure*}

In Fig.~\ref{fig:fred2d}, we present the \fred\ and
\fpass\ distributions in the $(\mhalo,\mgal)$ plane.  We pixelise the
$(\mhalo,\mgal)$ plane, compute \fred\ and \fpass\ in each pixel, and
indicate its value by the colour of the pixel.  For example, a red
colour indicates that galaxies are mostly red or passive within the
pixel, while a blue colour indicates that most of the galaxies are
blue/active. These diagrams reveal a number of interesting
features. Considering the diagram for central galaxies in the empirical
sample, we see that above a critical halo mass ($\mhalo \gtrsim 
10^{13}h^{-1}\msun$), nearly all galaxies are red and passive. For
intermediate halo masses $10^{11} \lesssim \mhalo \lesssim 10^{13}
h^{-1}\msun$, the structures show a complex dependence on both 
halo mass and stellar mass. In particular we note that above a ``critical''
stellar mass of $2-3~10^{10} h^{-2}\msun$, the majority of
galaxies are red and passive, regardless of their halo mass (though
such massive galaxies are not found in halos less massive than
$10^{12} h^{-1}\msun$).
Comparing with the models that showed good
qualitative behavior in terms of the binned quantities (S08 fiducial,
S08 halo-quenching, and dL06), we can see that the distribution of the
patterns in $(\mhalo,\mgal)$ space are quite different from the
observations --- in general, the structures show stronger vertical
divisions, indicating a stronger dependence on halo mass than on
galaxy mass. It is interesting to note that these three models look
much more similar to one another than any of the models does to the
empirical data. It is also interesting that in terms of \fred, the S08
fiducial and S08 halo-quenching models look very similar to one
another, but they look extremely different in the
\fpass\ diagram. This again illustrates that optical colours are not
an ideal probe of star formation quenching.

The observed satellites show an interesting striation, which is nearly
horizontal at the highest and lowest stellar masses, but somewhat
diagonal for intermediate masses. It appears that the majority of
satellite galaxies are red/passive if they are more massive than a few
times $10^{10} h^{-2}\msun$ (just as for the central population), and
are predominantly blue/active if they are less massive than $10^{10}
h^{-2}\msun$, regardless of their halo mass. For intermediate halo
masses, it seems that unlike for central galaxies, the critical mass
that marks the transition between mostly blue and mostly red galaxies
{\em is a function of halo mass}, and is lower for higher halo
masses. 
This suggests that the star formation activity in the most massive satellites is
regulated by the same processes that shape centrals, while lower mass
satellites are influenced by environmental processes such as tidal
forces or ram pressure stripping.

All the models fail quite miserably to reproduce the satellite
properties. In addition to simply predicting too high a fraction of
red satellites, none of the models shows the diagonal pattern of
contours in the \fred\ or \fpass\ diagrams. The few pixels with high
blue fractions in the models lie at high stellar mass for their halo
mass, which is the opposite from what is seen in the
empirical data. These are simply galaxies that were forming stars as
centrals and have become satellites very recently. 

Readers are referred to the Appendix for a discussion of the impact on our 
results of the modelling of dust extinction, our imposed selection criteria,
and possible biases in the procedure for assigning halo masses in the
SDSS group catalog.

\subsection{Connection with Morphology, Black Hole Formation and AGN Feedback}
\label{sec:conclusions}

\begin{figure}
\begin{center}
\includegraphics[width=8.5cm]{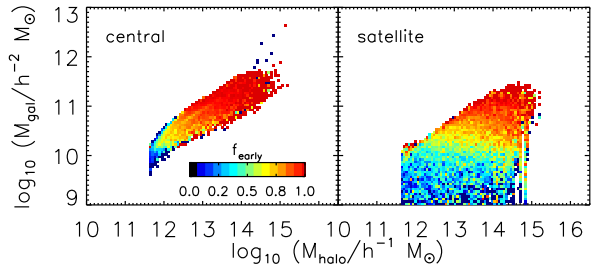}
\caption{ The fraction of early-type galaxies ($\rm{f_{early}}$) for
  centrals ({\it left}) and satellites ({\it right}) in the SDSS group
  catalog. Early type galaxies are defined as having a concentration
  index ($C$) greater than 2.6.  }
\label{fig:fearly}
\end{center}
\end{figure}

These results naturally beg the question: which physical process(es)
are responsible for imprinting this dependence of star formation
quenching on galaxy and halo mass? Although this is a complex question
that we will not be able to fully address in this paper, we attempt to
at least identify some promising hypotheses that can be pursued futher
in the future.

As already discussed by many authors \citep[e.g.][]{kauffmann03}, we
are suspicous that the correlation of star formation quenching with stellar 
mass may in fact be linked to the tendency of the {\em increasing bulge 
fraction} also with stellar mass and perhaps halo mass. Although we
have only rough morphological information for SDSS, we make use of a
standard cut in concentration index ($C\equiv r_{90}/r_{50}$) to
coarsely divide our sample into early and late type galaxies (early
types have $C>2.6$; \citealt{strateva01,shimasaku01}). Such
classification using the concentration index is subject to
contamination at roughly the 20\% level \citep{strateva01}. We then
plot $f_{\rm early}$ in the (\mhalo,\mgal) plane as
before\footnote{Because of the difficulty of mapping the morphological
  information available in the models to the observable concentration
  index, we do not show the model predictions. The qualitative trends
  in the models are similar.}, and show the results in
Fig.~\ref{fig:fearly}. We see a strikingly similar pattern to the one
seen when we plotted \fred\ and \fpass\ in this way in
Fig.~\ref{fig:fred2d}, indicating a very strong correlation between
red, passive, and early type galaxies. 

\begin{figure}
\begin{center}
\includegraphics[width=8cm]{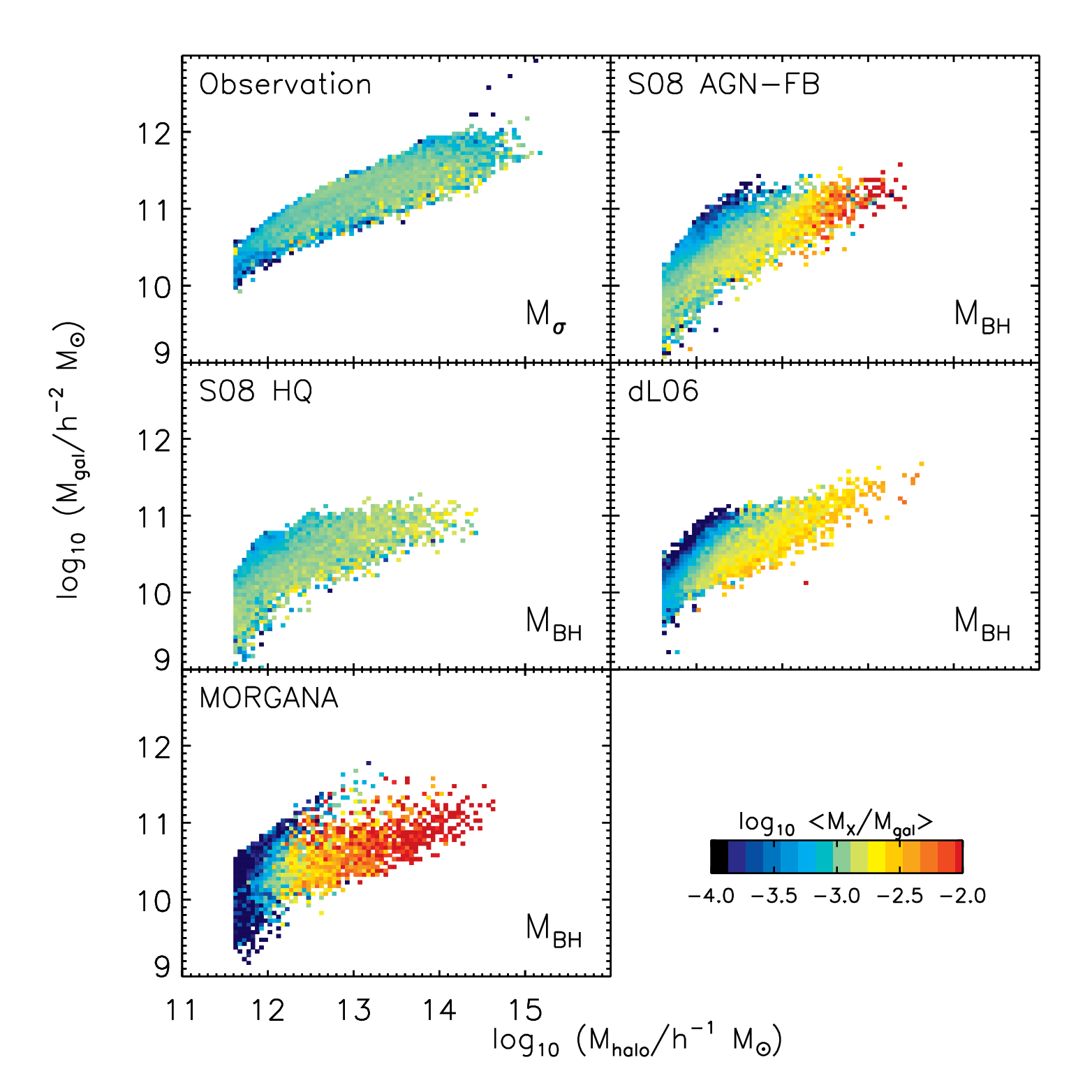}
\caption{Average black hole mass to galaxy stellar mass ratio
  $\left<M_{BH}/M_{gal}\right>$ for central galaxies, shown by the
  colour coding, as a function of \mhalo\ and \mgal, for four of the
  theoretical models.  For the S08 and dL06 models, note the
  similarity of the structures seen here to those seen in the plot of
  \fpass\ in Fig.~\protect\ref{fig:fred2d}. This suggests that for
  intermediate halo masses, \fpass\ is closely related to
  $\left<M_{BH}/M_{gal}\right>$ in the models.  We also show
  $\left<M_{\sigma}/M_{gal}\right>$ for the SDSS group catalogs, where
  $M_{\sigma}$, based on the SDSS measured velocity dispersion and the
  observed $M_{\rm BH}-\sigma$ relation, is used as a proxy for the
  average black hole mass.  Interestingly, the strong dependence of
  $\left<M_{\sigma}/M_{gal}\right>$ on halo mass is not visible.  }
\label{fig:mbhmgal}
\end{center}
\end{figure}

We test the connection of black hole mass with star formation quenching by plotting the
ratio of black hole mass to the total stellar mass of the galaxy
($\left<M_{BH}/M_{\rm gal}\right>$) in the same (\mhalo,\mgal) plane,
which we do in Fig.~\ref{fig:mbhmgal} (for central galaxies
only). Note that all three semi-analytic model codes reproduce the
empirical relation between bulge mass and black hole mass reasonably well
within the observational errors. In the S08, dL06, and {\sc morgana}
models, we see that the galaxies that have the {\em highest} stellar
mass to halo mass ratios have the {\em lowest} black hole mass to stellar mass
ratios. This is a reflection of the spread in halo merger histories at
fixed halo mass. In halos in which a massive black hole is formed relatively
early, the cooling flow is also shut off at an earlier time, halting
further galaxy growth except by mergers. Conversely, halos with
relatively small black holes for their mass will be able to continue to cool,
and the central galaxies will continue to form stars and remain
blue. This is seen to be the case in the ``blue ridge'' in
Fig.~\ref{fig:fred2d} (see panels h, k, and l). 
Note that neither the strong trend in
$\left<M_{BH}/M_{\rm gal}\right>$ nor the ``blue ridge'' is seen in the
S08 ``halo quenching'' model, in which quenching is regulated purely
by the halo mass.

In all three models with AGN feedback, there is also a trend with halo
mass in the sense that larger mass halos have larger black
hole-to-stellar mass ratios.  This is due to the dual modes of black
hole growth in the models. In the ``bright mode'', the growth of the
stellar bulge and the black hole are linked. Most of the black hole
growth in the models occurs via this bright mode of black hole
feeding. However, at late times, large mass halos can develop a hot
hydrostatic halo which is assumed to fuel the ``radio mode'' of black
hole growth. In these halos, the black hole can grow without any
associated star formation, leading to an increase in
$\left<M_{BH}/M_{\rm gal}\right>$. This is supported by the fact that
we do not see such a trend in the halo quenching model, which only
contains black hole growth via the bright mode.

We attempt to investigate whether this trend exists in the
empirical data, using the SDSS velocity dispersion $\sigma$ as a proxy
for black hole mass. We compute $M_{\sigma}$ from the $M_{BH}-\sigma_e$
relation using the empirical relation of \citet{gebhardt00}. We assign
a scatter to the black hole mass at a given $\sigma$ by choosing a uniform
random deviate over the 1$\sigma$ range quoted by \citet{gebhardt00}
(we obtain indistinguishable results when a Gaussian random deviate is
used). Interestingly, we see no evidence of a trend in
$\left<M_{BH}/M_{\rm gal}\right>$ with halo mass.

However, this result is not strongly conclusive, because it is not
known whether the relationship between $M_{BH}$ and $\sigma_e$ depends
on halo mass. However, these results are suggestive that there may be
too much black hole growth via the ``radio mode'' in the models,
perhaps indicating that other heating processes {\em not associated
  with black hole growth} may also be important in quenching cooling
flows.

We caution, as well, that although it is tempting to try to interpret
the ``fine features'' in these diagrams, the assignment of halo masses
in the current SDSS group catalogs is not very precise (see the
Appendix), and therefore only broad statistical trends should be taken
seriously. It is possible that in the future, if more precise
estimates of individual halo masses become available for large
samples, for example from gravitational lensing or X-rays, we may be
able to investigate these predicted trends in more detail.

\section{Discussion and Conclusions}

In this paper, we set out to investigate the significance of internal
galaxy properties vs. environment in shaping the star formation
history of galaxies, and to attempt to understand some of the physical
processes that might be at work. We made use of the observational
Group Catalog constructed from SDSS DR4 and the NYU-VAGC
\citep{yang07}.  The Group Catalog provides an estimate of the halo
mass for each group, which can be directly compared with semi-analytic
models. These halo mass estimates should provide a more unbiased probe
of global environment than measures based on local galaxy density
(e.g. distance to the $n$-th nearest neighbor)
\footnote{But note that some groups have started measuring
  galaxy number densities using search ellipsoids large
  enough to be representative of massive cluster halos 
  \citep[e.g.,][]{schawinski07a,yoon08}}.  

We have also used a sub-sample of SDSS with GALEX coverage to
estimate star formation rates \citep{salim07}, which should be a more
direct probe of the physics of star formation quenching than optical
colours. Besides, we make use of semi-analytic models from several
independent groups (S08, dL06, {\sc morgana}), and containing
different sets of recipes representing physical processes.

We first investigated the global distribution of colour vs. magnitude
and SSFR vs. stellar mass in the five models we considered, compared
with the empirical data. Although the different models showed some
differences in the details of their predictions for these quantities,
the models with some kind of quenching (either due to AGN feedback or
according to a critical halo mass) have similar qualitative features
and on the whole are a reasonable match to the empirical data --- not
surprisingly, as these observational quantities have been a target for
theoretical models for some time. We therefore found that it was
sensible to identify active or quenched galaxies according either to a
colour-magnitude criterion (the usual ``green valley''), or a similar
SSFR-\mgal\ criterion.

Next we investigated the stellar mass dependence of the quenched
fraction, based on optical colours (\fred), or on SSFR determined from
SDSS+GALEX photometry (\fpass). Here we largely confirmed and
reproduced results shown previously by other authors (though our
results in terms of SSFR from GALEX are new), namely that the fraction
of red/passive galaxies increases with stellar mass for both central
and satellite galaxies, and that the models with AGN feedback or halo
mass-based quenching reproduce this trend reasonably well for central
galaxies, but fail badly for satellites. All of the models produce too
high a fraction of red satellites and too flat a dependence of
quenching on stellar mass, which we term the {\em satellite
  over-quenching problem}.

We then investigated the joint dependence of quenched fraction
(\fred\ and \fpass) on galaxy mass and halo mass. First we
investigated \fred\ and \fpass\ as a function of halo mass, in
different stellar mass bins. A difficulty with this approach was that,
especially for central galaxies, there is quite a limited range of
stellar masses in halos of a given mass. 
Our analysis showed that, for central galaxies in
the observational group catalogs, the fraction of quenched galaxies
shows a strong dependence on halo mass at fixed stellar mass, but
also shows a strong dependence on stellar mass at fixed halo
mass. We were not able to determine which quantity is the primary
driver of quenching.
The S08 fiducial and dL06 models reproduced these trends fairly well. 
For observed satellite galaxies, there was a much stronger dependence on 
stellar mass visible in the \fred\ diagram (based on optical colours), 
and a weaker trend in the \fpass\ diagram (based on UV-derived SSFR). 
Once again, all models failed to reproduce the satellite properties, 
and even showed an inverted trend in \fpass\ with respect to the 
empirical data.

We also found it interesting to look at the pattern of \fred\ and
\fpass\ in terms of the two dimensional (\mhalo-\mgal) plane. 
This analysis revealed that the contours of
\fred\ and \fpass\ for central galaxies can be interpreted either as
a horizontal run or as a vertical run, again due to the halo mass --
central galaxy mass degeneracy.  The models showed quite a
different pattern in this space, and tended to show stronger vertical
boundaries, indicating a stronger dependence on halo mass.  For both
the empirical data and the models, these diagrams demonstrate the
complexity of the interplay between halo mass and stellar mass, and
are a promising tool for posing stringent tests on physical recipes in
galaxy formation models.  However, we caution that the estimates of
halo mass in the SDSS group catalogs are statistical in nature, and
this may introduce distortions into these distributions (see
Appendix).

We attempted to probe the physical origin of these results by
exploring additional correlations, such as the fraction of
morphologically early type (spheroid dominated) galaxies in the
observed sample, \fearly. We found a strikingly similar pattern for
\fearly\ in the (\mhalo,\mgal) plane as we did for \fred\ and \fpass,
suggesting that these quantities are tightly linked in some way. One
natural possibility is that the bulge mass is correlated with the mass
of a supermassive black hole, and that the black hole mass in turn
controls the quenching of star formation.  Intriguingly, we found that
in the models with black hole-regulated AGN heating (S08 fiducial and
dL06), the galaxies that were most likely to be blue, actively star
forming, and disk dominated, were expected to be those with the
smallest black hole for their mass. In addition, we saw a trend with
halo mass, in the sense that central galaxies in larger mass halos are
able to grow black holes more efficiently (the ratio of black hole mass
to stellar mass is larger). These trends were much weaker or absent in
models in which the black hole is not involved in regulating cooling
(such as the ``halo quenching'' model). We do not have direct
estimates of black hole masses for large samples of galaxies, but we
used the measured SDSS velocity dispersion $\sigma_e$ and the observed
$M_{BH}-\sigma_e$ relation to obtain estimates of a black hole mass
proxy, $M_{\sigma}$. We do not see a strong trend in
$M_{\sigma}/\mgal$ for the SDSS group catalog, indicating either that
this method for estimating the empirical black hole masses is too
crude, or that the dependence of black hole mass on halo mass in the
models is too strong.

Although the model predictions for the distribution of \fred\ and
\fpass\ in the (\mhalo, \mgal) plane do not match the observational
results in detail, we conclude that the observational data
are consistent with the basic qualitative picture
presented by the models in which cooling is regulated by AGN feedback
at least for central galaxies. In these models, the suppression of
cooling and the quenching of star formation depend on two factors: the
presence of a quasi-hydrostatic hot halo (strongly correlated with
halo mass), and the mass of the supermassive black hole (strongly
correlated with galaxy mass). This picture is also strongly supported
by the recent work of \citet{pasquali08}, which directly explored the
dependence of ``radio mode'' and ``bright mode'' AGN activity on halo
mass and stellar mass in these same SDSS group catalogs.  The
empirical data do not seem to support models in which the process that
suppresses cooling is solely a function of halo mass
\citep[e.g.,][]{dekel08,khochfar08}, although it will be important to
explore the explicit predictions of alternate heating mechanisms (such
as heating by clumps or infalling satellites) in detail.

The {\sc morgana} model suffers from the largest disagreement with the
observations. The treatment of the triggering of ``radio mode''
accretion in {\sc morgana} is significantly different than in the
other two models, requiring that star formation is inevitably
associated with the triggering of radio mode accretion. Although the
radio mode feedback mechanism adopted in {\sc morgana} is largely able
to solve the ``overcooling problem'' in terms of reproducing the
galaxy stellar mass or luminosity function, this star formation makes
many massive galaxies too blue. Our analysis places tight constraints
on the links between star formation and AGN activity at late times,
and highlights our current lack of understanding of the details of the
processes that regulate both kinds of activity.

For satellite galaxies, the empirical diagrams suggested that
quenching depends on {\em both} stellar mass and halo mass, such that the
``critical stellar mass'' that divides active from passive (blue from
red) galaxies decreases with increasing halo mass.  None of the models
was successful in predicting this trend.  
Including tidal destruction of satellites, as was
done in the S08 models, improves the agreement with the data because
satellites that have been orbiting for a long time within the host
halo (which tend to be red and passive) are destroyed. However, it
seems that tidal destruction cannot provide a full solution to the
problem.

The problems with
over-quenching of satellites in semi-analytic models have been
demonstrated before \citep[e.g.][]{weinmann06b} and are likely due to
the assumption applied in nearly all SAMs that the hot gas halo, which
is the source of new cooling gas, is stripped off instantly when a
galaxy becomes a satellite in a larger halo (sometimes called
``strangulation''). Therefore satellites fairly quickly consume their
remaining cold gas reservoirs and become red and passive
\citep[e.g.][]{crowl06}. However, recent hydrodynamic simulations have
found that the hot gas halos of satellites are not stripped instantly
\citep{kawata08,mccarthy08}.  Recently, several authors
\citep{kang08,font08} have proposed improved recipes for the treatment
of cooling onto satellites, which produce better results for the
predictions of satellite colours. Clearly, our analysis should be
repeated with one of these improved treatments implemented in our
models. However, it is probably also important to properly treat the
effects of ram pressure stripping on both the satellites' hot halo and
the cold gas in the galaxy \citep{lanzoni05,quilis00,okamoto03}.  This
will clearly be an important area for improvements to the modelling
and further investigations.

\section*{Acknowledgments}
\begin{small}
The empirical data derived from the Sloan Digital Sky Survey and the
Galaxy Evolution Explorer (GALEX) observations played a critical role
in this project.  We warmly thank G. de Lucia and G. Lemson for help
with the Millennium Catalogs and database server, and Sadegh Khochfar
and Marc Sarzi for numerous insightful discussions.  We are also grateful
to the referee for several important clarifications. SKY acknowledges
support from the Basic Research Program of the Korea Science and
Engineering Foundation (R01-2006-000-10716-0).  TK is grateful for the
hospitality of the Max-Planck-Institut f\"ur Astronomie in Heidelberg
during his visit.  FF and PM thank Laura Silva for help in the use of
GRASIL.  Some of the calculations were carried out at the PIA cluster
of the Max-Planck-Institut f\"ur Astronomie at the Rechenzentrum
Garching.

\end{small}

\newpage

\begin{figure*}
\begin{center}
\includegraphics[width=8cm]{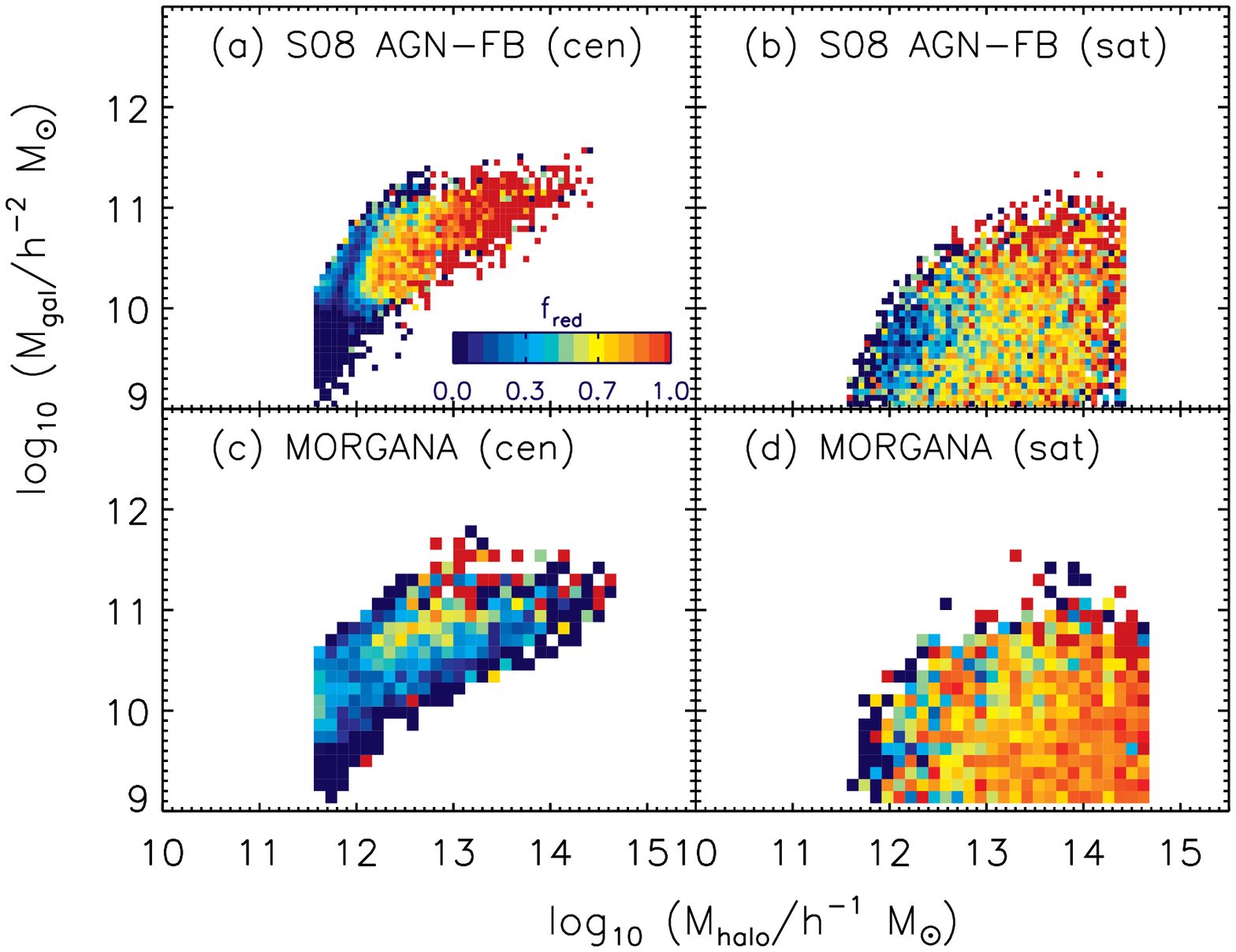}
\includegraphics[width=8cm]{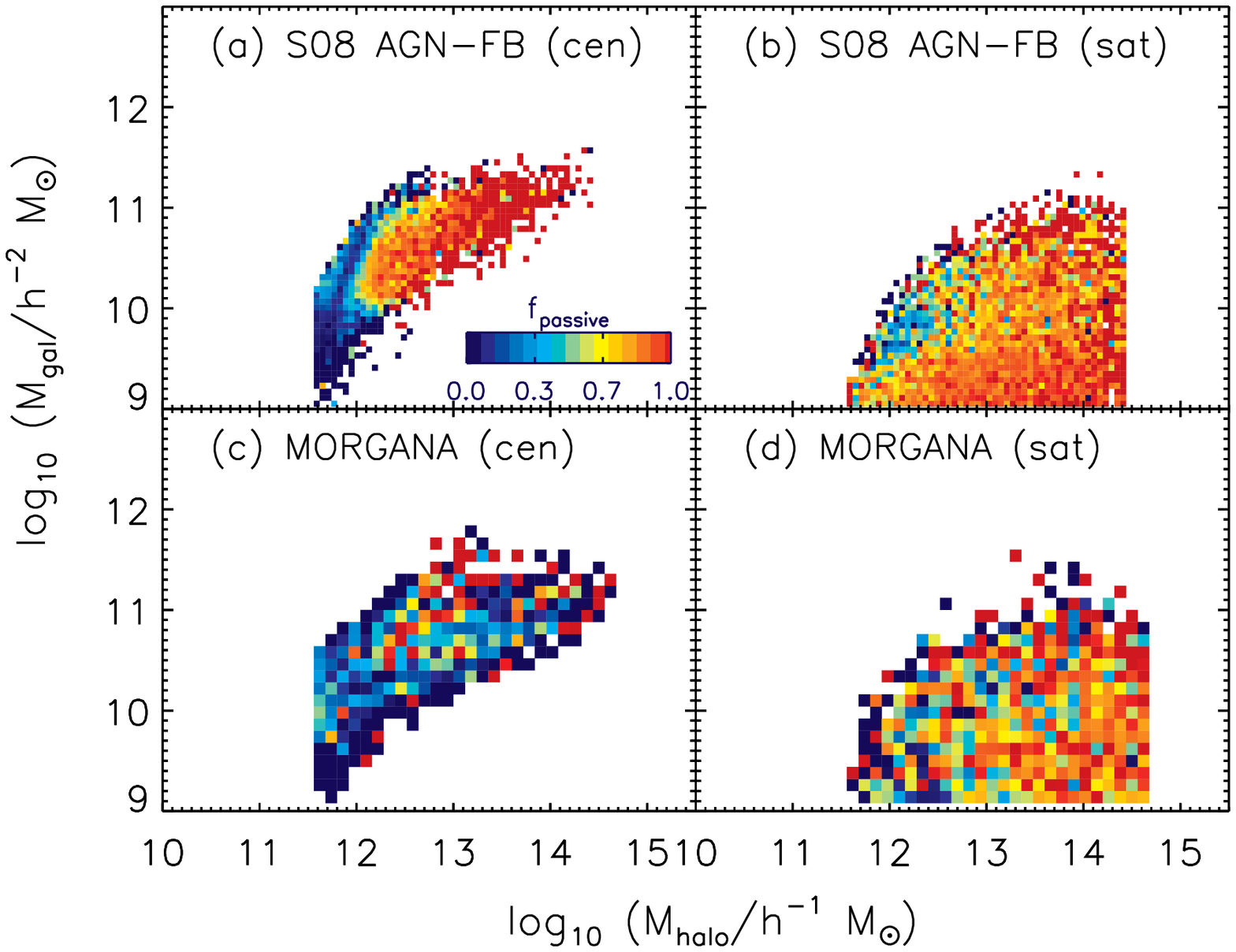}
\caption{The fraction of "red" (\fred) (left panels) and "passive"
  galaxies ($f_{\rm passive}$) (right panels) in models without dust
  corrections. The colour scale is as in Fig.~\ref{fig:fred2d}. We
  present the results for central galaxies (left) and satellite
  (right) galaxies seperately.  We see that the details of the
  \fred\ distribution are quite sensitive to the dust correction,
  whereas $f_{\rm passive}$ is not noticably affected by the dust
  correction. We also note that the dust-free \fred\ results appear
  more similar to the \fpass\ results, which presumably probe the
  physical properties of galaxies more directly. }
\label{fig:fred2d_nodust}
\end{center}
\end{figure*}

\begin{figure*}
\begin{center}
\includegraphics[width=8cm]{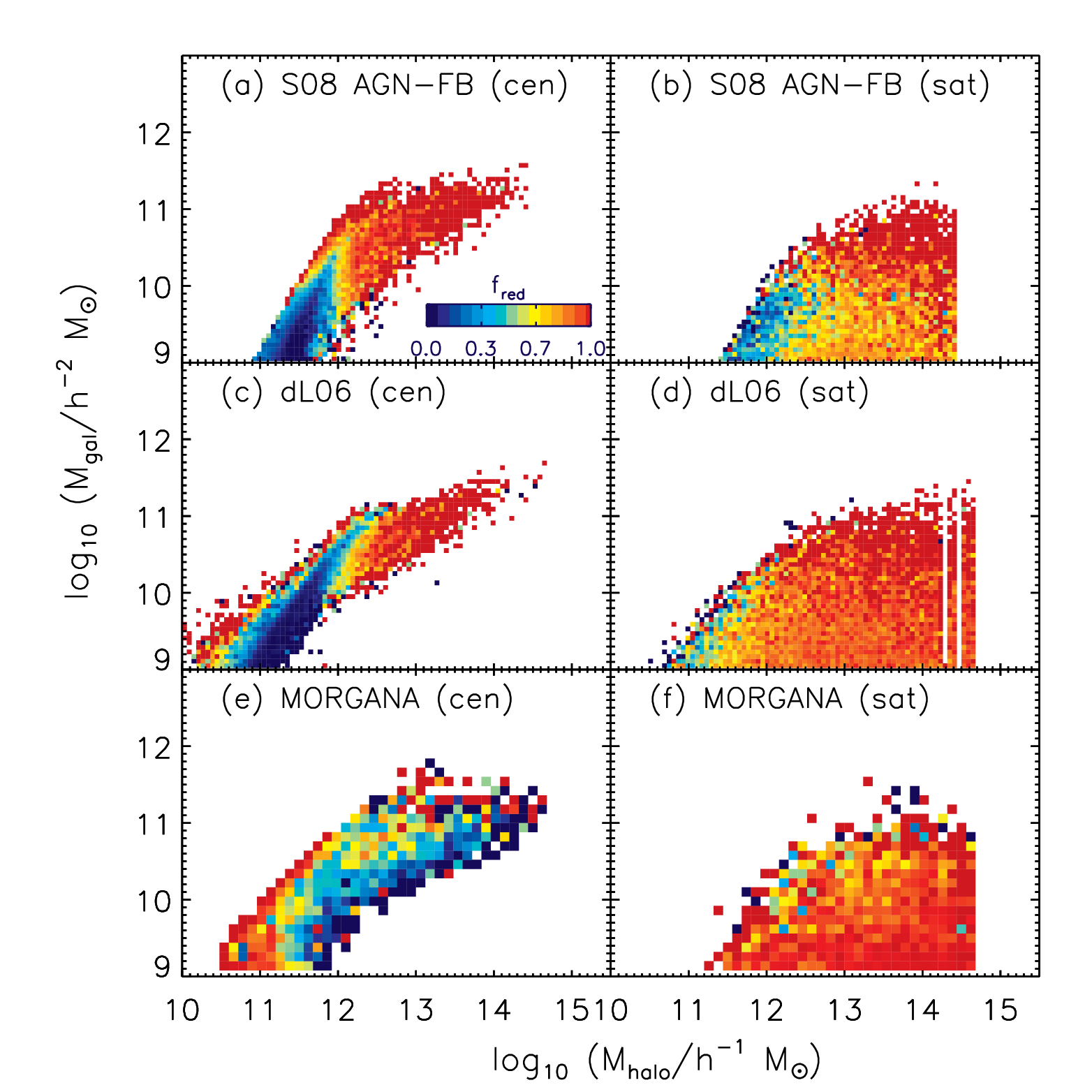}
\includegraphics[width=8cm]{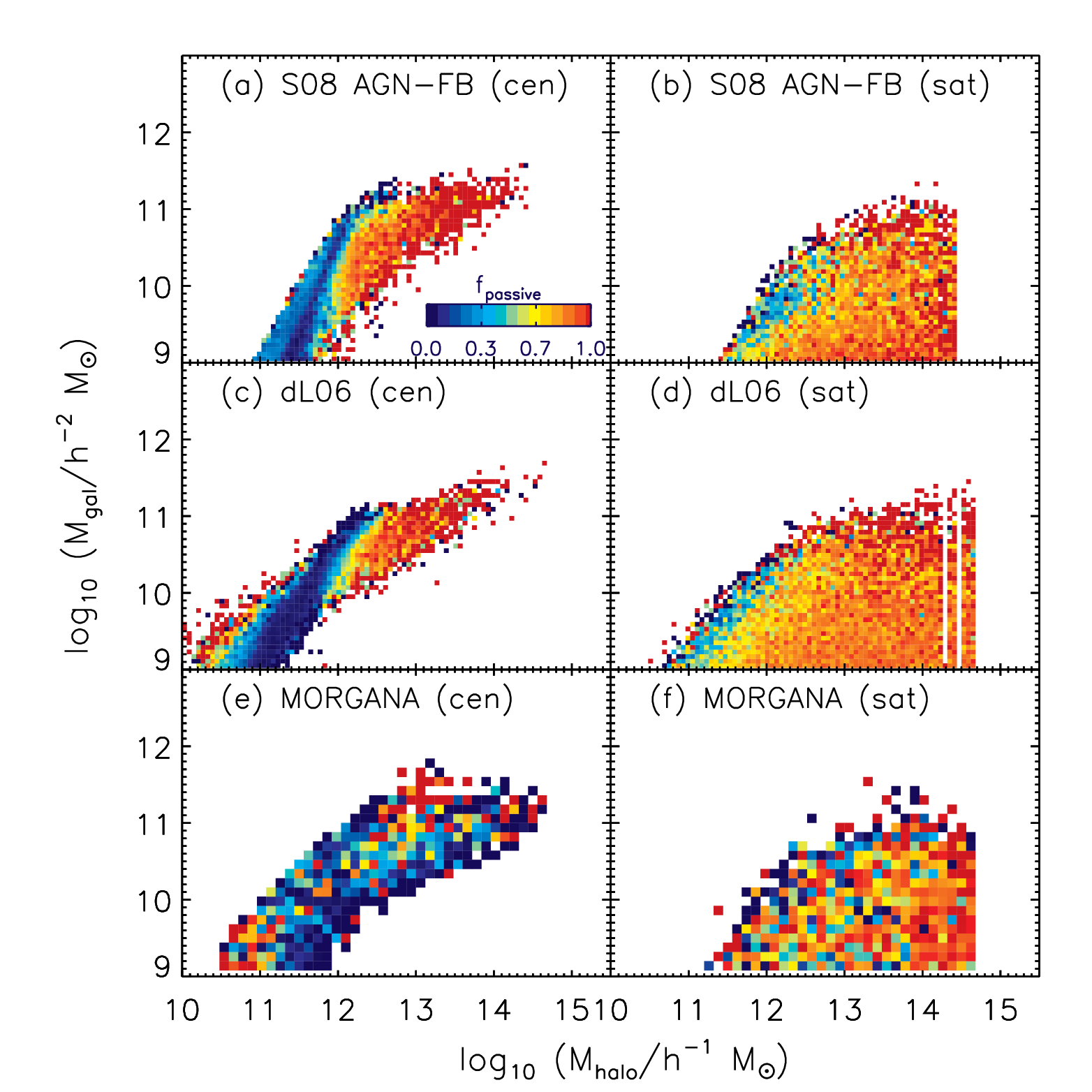}
\caption{ The fraction of ``red'' galaxies (\fred) (left panels) and of
  ``passive'' galaxies ($f_{\rm passive}$) (right panels) without
  selection criteria. The colour scale is as in
  Fig.~\ref{fig:fred2d}. We present the results for central
  galaxies (left) and satellite (right) galaxies seperately.  It can
  be inferred that our main results do not depend on our selection
  criteria.  }
\label{fig:fred2d_noselec}
\end{center}
\end{figure*}

\section*{Appendix : The effect of dust, selection criteria, and group catalog halo mass estimates}

All of our models include a treatment of dust extinction, which
affects both the colours and magnitudes of the model galaxies. In
addition, as mentioned in \S 4, for our main analysis we have applied
selection criteria to the models to mimic those that we believe to be
present in the SDSS group catalog. We include only galaxies with
apparent $r$-band magnitude brighter than 17.77 mags, and we also
excluded halos which do not contain any galaxy member brighter than
$^{0.1}M_r \leq -19.5 + 5\log h$. In this Appendix we provide the
results for the 2-d distributions of \fred\ and \fpass\ for the
theoretical models without dust extinction and without selection
criteria applied. In addition, we test for possible biases that may
arise from the approach used to assign halo masses to the groups in
the SDSS group catalogs by applying this method to the model mock
catalogs.

In Fig.~\ref{fig:fred2d_nodust} we show \fred\ and \fpass\ without any
`dust corrections applied to the models. Since we do not have magnitude
information without dust corrections for the dL06 models, only the S08
fiducial and {\sc morgana} models are shown. Interestingly, the
results now look much more similar to the results seen before for
\fpass. In particular, the blue ridge which was visible in the
\fpass\ diagrams corresponded to a {\em red} ridge in the
\fred\ diagram (Fig.~\ref{fig:fred2d}). The blue ridge is now visible
in \fred\ as well, indicating that these galaxies are actually
actively star forming, and were predicted to be red only because of
dust extinction. In the case of \fpass, the dust correction only
affects the galaxy selection and causes a negligible change in the
diagrams. The treatment of dust extinction is one of the most
uncertain aspects of the modelling, and this highlights the advantage
of using intrinsic physical quantities extracted from the
observations.

Fig.~\ref{fig:fred2d_noselec} shows the model predictions with no
selection effects applied. Comparing this with Fig.~\ref{fig:fred2d}, we
see that the results appear unchanged above a halo mass $\log
\mhalo/h^{-1}\msun \ge 11.6$ and a stellar mass $\mgal \ge 10^{9}
\msun$. This is reassuring in the context of our present
analysis. However, we can also see that there is interesting predicted
behavior at lower halo and galaxy masses than we can currently probe,
and also interesting differences between the models at these
masses. This suggests that it would be extremely useful to obtain
similar data that are complete to fainter levels, so that we could
probe lower mass galaxies and lower mass halos.

\begin{figure}
\begin{center}
\includegraphics[width=8cm]{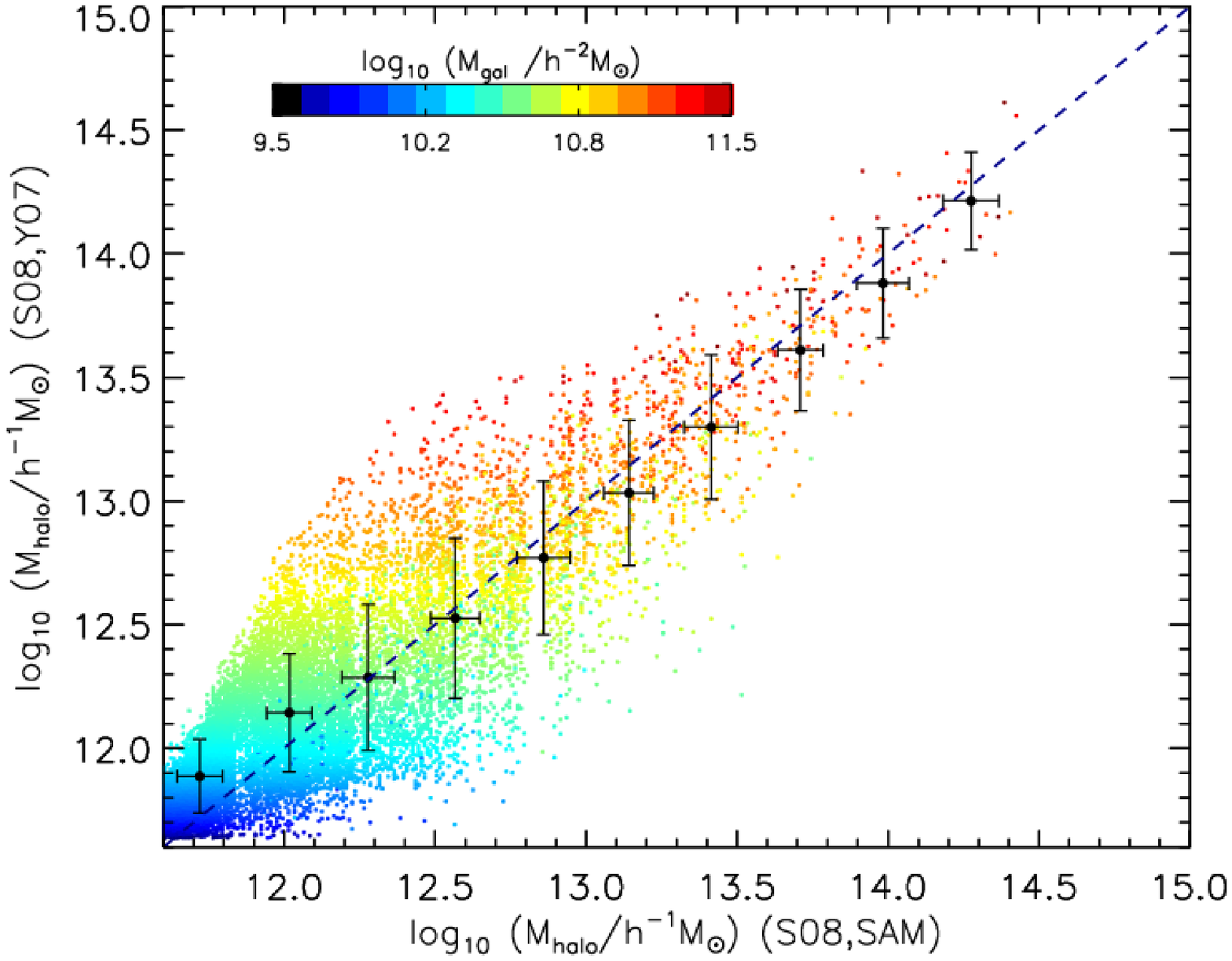}
\caption{The true halo mass in the fiducial semi-analytic model of
  S08, compared with the halo mass estimate based on the total stellar
  mass in the halo, using a procedure similar to that of Y07 for the
  SDSS-based group catalog. The colour-scale indicates the stellar mass
  of the central galaxy in each halo, as shown by the key on the
  figure. Although the mean halo mass is reproduced fairly well, there
  is quite a large scatter in the true halo mass at a given estimated
  halo mass.}
\label{fig:mhalocomp}
\end{center}
\end{figure}

Finally, we investigate the procedure used to assign halo masses to
the groups that are identified in the SDSS group catalog. For the
results presented here, stellar masses for several models are obtained 
from \citet{bell03} like Y07, and halo masses are assigned based on the
``characteristic'' stellar mass of the group, where the characteristic
stellar mass is defined as the total stellar
mass contributed by galaxies with $^{0.1}M_r \leq -19.5 + 5\log h$. 
Halo masses are then assigned by matching the rank-ordered list of
group characteristic stellar masses with a rank-ordered list of dark
matter halo masses from a theoretical estimate of the DM halo mass
function, assuming a monotonic mapping between the characteristic
stellar mass and DM halo mass (see \S\ref{sec:data:group} and Y07). We
apply this procedure to the halos in the mock catalog produced with
the S08 fiducial semi-analytic model, and show the comparison between
the true and estimated halo mass in Fig.~\ref{fig:mhalocomp}. We see
that although the mean halo mass is estimated fairly accurately, there
is a large scatter in true halo mass at a given value of the estimated
halo mass.

\begin{figure}
\begin{center}
\includegraphics[width=8cm]{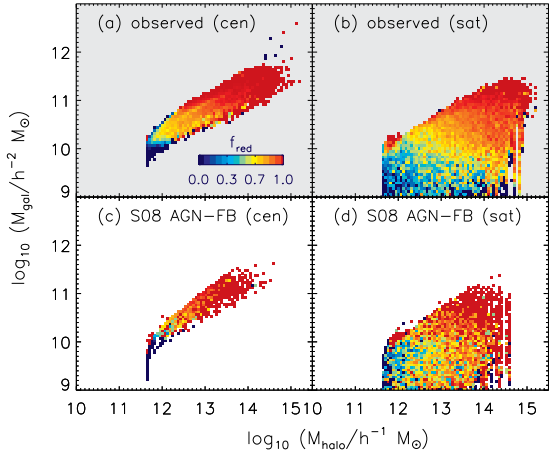}
\caption{The fraction of ``red'' galaxies (\fred) as a function of halo
  mass and stellar mass, where halo masses have been assigned in the
  semi-analytic models using an approach similar to that used in the
  SDSS group catalogs. The colour scale is as in
  Fig.~\ref{fig:fred2d}. We present the results for central
  galaxies (left) and satellite (right) galaxies seperately. We see
  that the procedure used to assign halo masses in the SDSS group
  catalogs reduces the scatter in \mgal\ at fixed halo mass, and
  washes out many of the detailed features of the 2d distribution that
  are visible in the raw model predictions. }
\label{fig:fred2d_yang}
\end{center}
\end{figure}

In Fig.~\ref{fig:fred2d_yang}, we show again the distribution of
\fred\ with halo mass and stellar mass, now using the Y07-like halo
mass estimates for the semi-analytic models instead of the true halo
masses. We see that the procedure artificially reduces the scatter in
\mgal\ at fixed halo mass. 
This can be attributed to two effects.  First, the halo mass estimates
in the Y07 group catalogue are based on the total stellar mass in the
halo, which strongly correlates with the stellar mass of the central
galaxy.  In addition, the Y07 group catalogue includes corrections for
various incompleteness effects in the SDSS (the factor {\cal C} in
Eqs. 3--4 in Y07), which creates scatter that is not visible in the
mock catalogues.  It should also be noted that the halo-mass
estimating algorithm of Y07 washes out many of the detailed features
that are visible in the model predictions.  Hence we should not make
too much of the detailed features in the empirical
\mgal-\mhalo\ diagrams but should focus on the mean trends instead.

\end{document}